# Large zero-bias peaks in InSb-Al hybrid semiconductor-superconductor nanowire devices

Hao Zhang*,[1, 2, 3] Michiel W.A. de Moor*,[1, 2] Jouri D.S. Bommer*,[1, 2] Di Xu,[1, 2] Guanzhong Wang,[1, 2] Nick van Loo,[1, 2] Chun-Xiao Liu,[1, 2, 4] Sasa Gazibegovic,[5] John A. Logan,[6] Diana Car,[5] Roy L. M. Op het Veld,[5] Petrus J. van Veldhoven,[5] Sebastian Koelling[a],[5] Marcel A. Verheijen,[5] Mihir Pendharkar,[7] Daniel J. Pennachio,[6] Borzoyeh Shojaei,[6, 8] Joon Sue Lee[b],[8] Chris J. Palmstrøm,[6, 7, 8] Erik P.A.M. Bakkers,[5] S. Das Sarma,[4] Leo P. Kouwenhoven[1, 2, 9†]

[1]QuTech, Delft University of Technology, 2600 GA Delft, The Netherlands
[2]Kavli Institute of Nanoscience, Delft University of Technology, 2600 GA Delft, The Netherlands
[3]State Key Laboratory of Low Dimensional Quantum Physics, Department of Physics, Tsinghua University, Beijing 100084, China
[4]Condensed Matter Theory Center and Joint Quantum Institute, Department of Physics, University of Maryland, College Park, Maryland 20742, USA
[5]Department of Applied Physics, Eindhoven University of Technology, 5600 MB Eindhoven, The Netherlands
[6]Materials Engineering, University of California Santa Barbara, CA 93106, USA
[7]Electrical and Computer Engineering, University of California Santa Barbara, Santa Barbara CA 93106, USA
[8]California NanoSystems Institute, University of California Santa Barbara, Santa Barbara, California 93106, USA
[9]Microsoft Quantum Lab Delft, 2600 GA Delft, The Netherlands

**We report electron transport studies on InSb-Al hybrid semiconductor-superconductor nanowire devices. Tunnelling spectroscopy is used to measure the evolution of subgap states while varying magnetic field and voltages applied to various nearby gates. At magnetic fields between 0.7 and 0.9 T, the differential conductance contains large zero bias peaks (ZBPs) whose height reaches values on the order $2e^2/h$. We investigate these ZBPs for large ranges of gate voltages in different devices. We discuss possible interpretations in terms of disorder-induced subgap states, Andreev bound states and Majorana zero modes.**

---

* These authors contributed equally to this work
† Leo.Kouwenhoven@microsoft.com
[a] Current address: Department of Engineering Physics, Polytechnique Montreal, 2500 Chemin de Polytechnique, Montreal, Canada
[b] Current address: Department of Physics and Astronomy, University of Tennessee, Knoxville, TN, 37996, USA

**Introduction**

Tunnelling spectroscopy in electrical transport is an important tool to measure the presence of states at energies smaller than the gap of bulk superconductors. Such subgap states exist, for instance, near magnetic impurities, inside vortices and at interfaces with other materials[1-3]. In hybrid combinations of a superconductor with a semiconductor, subgap states can have properties from both materials such as the particle-hole symmetry from the superconductor and the spin-related properties from the semiconductor. In confined regions, this leads to a variety of subgap states generally known as Andreev bound states (ABS)[1-3]. A magnetic field $B$ lifts the spin degeneracy of the ABS, turning them into non-degenerate, single-fermionic states. Tunnelling spectroscopy allows to measure the energy evolution as a function of $B$, enabling to identify the underlying quantum properties of individual ABS.

Recent interest has focussed on the possibility to induce a topological phase in a hybrid of a superconductor with a one-dimensional semiconducting nanowire[4-7]. A bulk-boundary correspondence dictates that the topological phase in the bulk comes together with localized boundary states at zero energy. These states are 'half-fermionic' with the two halves spatially separated at the two ends of the nanowire. Each half, if the two are sufficiently far from each other, is called a Majorana Zero Mode (MZM) and interestingly obeys non-Abelian statistics[8]. In the past decade there has been much theoretical and experimental interest to identify and study MZMs, with tunnelling spectroscopy as the primary tool [9-15].

The difficulty is, however, that interfaces, edges and nanowire-ends are often locations where subgap states arise from all kinds of non-topological origins, including electrostatic accumulation, disorder, and smoothly varying potentials. The challenge is to distinguish MZMs from these other types of localized states. Such a distinction may come from some of the peculiar signatures of MZMs, most prominently the requirement of zero energy independent of parameter values like $B$ or a gate voltage, $V_g$, that tunes the electron density. In tunnelling spectroscopy this shows up as a zero-bias peak (ZBP), which is called a "robust ZBP" when it remains at zero voltage-bias while varying $B$ or $V_g$ over significant ranges. The topological MZM interpretation also requires this ZBP to be quantized (at zero temperature) at a constant value of $2e^2/h$ with variations in systems parameters, i.e., the ZBP should be robust both in sticking to zero voltage and having a quantized value of $2e^2/h$ as magnetic field and various gate voltages are varied[16-18]. Such robust quantized ZBPs have not yet been observed experimentally. Note that finite temperature, finite tunnelling amplitude, finite wire length, dissipation and disorder could all affect the height of a ZBP[3,7].

A robust ZBP is, however, only a necessary property of topological MZMs but not sufficient to serve as an unambiguous proof of its existence. Numerical studies have shown that various shapes of electrostatic potentials near the end of a wire can also induce states at zero energy, without the need of a topological phase in the bulk of the nanowire[19-21]. Such states are denoted as 'trivial' (as opposed to 'topological') and are part of the ABS family. Since an ABS is a single-fermionic state and a MZM is half-fermionic, this gives rise to a distinguishing feature. Assuming zero temperature, then in the generic ABS case the height of the ZBP can reach a maximum value of $4e^2/h$ while for a MZM the ZBP cannot go above half that value, $2e^2/h$[17,18,22].

Numerical studies have also shown that trivial end-states (ABS) can be remarkably fixed to zero energy over extended regions in $B$ and $V_g$[21]. These zero-energy trivial states can be mathematically decomposed into two half-fermionic states. The difference with MZMs is that in this case, there is no topological phase in the bulk and the two half-fermionic states have strongly overlapping wavefunctions localized near the same wire-end. Depending on parameter values such as temperature and tunnel coupling to the spectroscopy lead, each half-fermionic state can contribute to the height of the ZBP by an amount between 0 and $2e^2/h$, and thus together yield a ZBP height between 0 and $4e^2/h$.

In fact, more detailed numerical studies[23-25] have shown that for parameter values relevant for the existing devices, one half-fermionic state of the ABS may occasionally couple much stronger to the lead than the other. At zero temperature the maximum contribution of just one half-fermionic state results in a ZBP height of $2e^2/h$, and without detailed knowledge of the system, it is indistinguishable from the signature of an MZM. This seriously complicates the desired, unambiguous distinction between an ABS and a MZM because they both may lead to an apparent ZBP height at $2e^2/h$ for the tunnelling conductance from one end.

The two half-fermionic states created from the decomposition of a trivial ABS have been denoted as 'partially separated ABS' (ps-ABS) to indicate some spatial separation of the two corresponding wavefunctions but only over a partial distance of the full wire length[23]. Alternatively, they have been called 'quasi-Majoranas' emphasizing that, despite the absence of a topological phase, these spatially-overlapping states are self-adjoint and individually obey non-Abelian statistics[24], even though they are not isolated.

An additional complication is that in the presence of a magnetic field, disorder can also induce states at energies smaller than the superconducting gap, including at zero energy[26]. These disorder-induced states may lead to ZBPs with conductance values anywhere between 0 and $4e^2/h$, and thus including $2e^2/h$. Disorder-induced ZBPs have a different dependence to variations in magnetic field and gate voltages[27,28], but it may require microscopic knowledge of the device for a clear distinction. Thus, ZBPs could arise from topological MZM, ps-ABS (or equivalently quasi-MZM) localized near the ends, or from disorder and are therefore by themselves insufficient to distinguish experimentally the different scenarios. We, moreover, note that each of these scenarios is expected to occur in very small parameter ranges. To find experimental proof for any of these scenarios requires a detailed search including fine tuning of parameters.

The majority of experimental papers[9-15] have reported ZBP heights much lower than $2e^2/h$, predominantly due to finite temperature and other broadening effects such as dissipation[29]. Tunnelling spectroscopy on nanowires defined in a two-dimensional electron gas have shown ZBP heights near $2e^2/h$[30]. We have previously reported tunnelling spectroscopy on InSb nanowires covered by Al, indicating ZBP heights with values close to $2e^2/h$ despite varying $B$ and $V_g$ parameters[31]. We reported this as an observation of conductance quantization serving as evidence for Majorana states[31]. After several issues were brought to our attention, we re-examined the data set as well as the technical details of our experimental setup. During this process, discrepancies with the original publication arose which led us to conclude that we can no longer claim a robust quantization, leading to a full retraction of the publication in Ref. 31.

In this manuscript we report the re-analysed data of Ref. 31, together with previously unpublished data from the same set of measurements. We first describe the experiments and present the data. At the end of this manuscript, we discuss the implications in the context of the various signatures for disorder-induced subgap states, ABS, and MZM.

Our work, presented here, along with the recent theoretical developments involving quasi-Majoranas and disorder, should serve as a strong cautionary message for all Majorana experiments in all platforms, clearly emphasizing that the experimental observations of zero-bias conductance peaks, no matter how compelling, should be considered only as necessary and by no means sufficient conditions for the existence of topological MZMs.

**Experiment**

Fig. 1 shows a scanning electron micrograph of device A. An InSb nanowire is covered by a thin superconducting Al shell. The nanowire is contacted by normal metal contacts, which are used to apply a bias voltage $V$ and drain a current $I$ to ground. The 'tunnel gates' are used to control the transmission through the uncovered segment between the left electrical contact and the wire section covered with the Al shell. A voltage $V_{TG}$ is applied to both tunnel gates simultaneously. The electron density in the nanowire segment covered by Al can be tuned by applying a voltage $V_{SG}$ to both of the 'super gates'. The doped substrate can be used as a global backgate. Unless otherwise noted, the backgate is grounded, i.e. $V_{BG} = 0$.

The device is mounted in a dilution refrigerator with a base temperature of about 20 mK, and a three-axis vector magnet which is used to align a magnetic field $B$ along the nanowire axis. We measure the differential conductance, $dI/dV$, through the device by applying a small ac excitation voltage and measuring the resulting current response using a lock-in amplifier. An extensive description of the measurement circuit is provided in Appendix B.

In Fig. 2 we show the differential conductance, $dI/dV$, versus $V$ and $B$ for different values of $V_{TG}$ with $V_{SG}$ fixed. At $B = 0$, we resolve the coherence peaks of the superconducting gap at $V = \pm 0.21$ mV (see also green curve in panel (e)). As $B$ increases, two levels detach from the gap edge and cross at zero bias, forming a ZBP. This is illustrated by the orange line cut in Fig. 2(e), showing a ZBP which reaches a value of ~$2e^2/h$ at 0.8 T. The evolution of the zero-bias conductance versus $B$ for different values of $V_{TG}$ is shown in Fig. 2(d). While for $V_{TG} = -7.74$ V the ZBP reaches a maximum height close to $2e^2/h$, this maximum can be decreased (blue curve) or increased (black curve) by changing $V_{TG}$. We also note that the dependence on $V_{TG}$ is non-monotonic, which will be further investigated in the next section. Due to uncertainties in some of the parameters of our setup, as well as the calibration procedure detailed in Appendix B, the conductance we extract has a typical error margin of a few percent. We illustrate the effect of this in Fig. 2(f) by zooming in on the section indicated by the dashed rectangle in Fig. 2(d). The extracted conductance is shown as a coloured line, with the error margin shown as a dark (1σ) and light (2σ) shading in the same colour. While a value of $2e^2/h$ for the maximum conductance is consistent with the error margin for $V_{TG} = -7.74$ V, it is outside the 2σ-margin for $V_{TG} = -7.8, -7.92$ V.

Next, we fix $B$ = 0.8 T and investigate the evolution of the peak height as a function of $V_{TG}$, shown in Fig. 3(a). The ZBP is present between $V_{TG}$ = -7.95 and $V_{TG}$ = -7.5 V, while for more negative values the peak splits away from zero energy. Near $V_{TG}$ = -7.6 V, there may be a sign of peak splitting. We also observe sudden changes in conductance which are likely due to charge rearrangements in the electrostatic environment of the nanowire. These instabilities, referred to as "charge jumps", are commonly observed in our devices. This is most likely due to the constraints placed on the fabrication process by the limited thermal budget[32], as too high temperatures will lead to intermixing between InSb and Al. This restricts our fabrication to room temperature processes which is not sufficient to remove all resist residues (visible in Fig. 1) and these can serve as charge traps. While some charge jumps occur stochastically and do not repeat upon remeasuring the same parameter range, other are reproducible. Charge jumps and hysteresis can lead to variation in the measured conductance values at nominally equal parameter settings (measurement history is specified in Appendix F). Additional discussion on charge jumps is provided in Appendix D.

In some cases, a charge jump seems to reset the electrostatic potential on the tunnel gate, causing conductance features to repeat after the jump. This is, for instance, visible in Fig. 3(a) at $V_{TG}$ = -7.97 V (indicated by the white arrow, see also Appendix D). Such charge jumps make the relation between tunnel barrier height and $V_{TG}$ non-monotonic. We can, however, compare the height of the ZBP to the out-of-gap conductance to investigate the effect of the tunnel barrier height on the ZBP.

In the range $V_{TG}$ = -7.93 V to -7.6 V (a significant charge jump occurred at -7.6 V), the zero-bias conductance fluctuates near $2e^2/h$, with a mean of 1.05 and a standard deviation of 0.05 in the unit of $2e^2/h$. (Fig. 3b). We consider this to be an example of "plateau-like" behaviour (see also Appendix E). In the same range, the out-of-gap conductance increases significantly, as shown in panel (c) by the line traces taken at $V$ = -0.2 mV (black) and $V$ = 0.2 mV (blue). We take the average conductance for $|V| \geq 0.2$ mV, $G_N$ (shown in green in Fig. 3(c)), as a proxy for the transmission to the lead. While the relative fluctuation in the ZBP height is 5%, $G_N$ increases by almost 40% over the same range, and generally increases with gate voltage. This indicates that ZBP-height does not simply scale with the transmission coupling to the lead.

We investigate this further in Fig. 4. In panel (a) we show linecuts taken from Fig. 3(a), where we have indicated the linecuts with a ZBP height close to $2e^2/h$ in red. We use a Lorentzian function, thermally broadened by 20 mK, to fit the peaks and extract their height and width (see Appendix G for details). Examples of these fits are shown in Fig. 4(c)-(e), together with the calculated $G_N$. This allows to plot the peak height and width as a function of $G_N$ in Fig. 4(b). The peak widths increase with $G_N$ and are larger than 50 µeV, indicating that the peak is mainly broadened by the coupling to the lead and not by temperature ($3.5k_BT$ = 15 µeV assuming an effective electron temperature of 50 mK). At the same time, there is no straightforward dependency of the peak height on $G_N$. Note that a small peak splitting may occur at high $G_N$.

We note that the measurements shown in Figs. 3 and 4 were selected after finetuning parameters in a dedicated search for quantized plateaus (see Appendix E for a discussion of our measurement methodology). We will discuss this dedicated search more in the discussion section.

The plateau behavior of the ZBP is striking but not convincingly present for all $B$ and $V_{SG}$ values where ZBPs are present. This can be seen in Fig. 5, which shows a scan like Fig. 3, but at a lower $B$-field of 0.7 T. Again, a ZBP is present over a sizeable $V_{TG}$ range, but in this case the peak height deviates from $2e^2/h$ with the maximum ZBP-height reaching ~1.5 × $2e^2/h$. The variation in the ZBP height is also significantly larger than the $G_N$ variation (Fig. 5c), different from the behaviour at 0.8 T in Fig. 3 (see Appendix E for quantified plateau criteria). Due to charge jumps and other instabilities, the effective potential can drift over time or change irreversibly after a large voltage swing. When returning to the same gate settings, the electrostatic potential is not necessarily the same. Fig. 2 was measured between 36 to 48 hours before Fig. 3 which was measured 30 hours before Fig. 5. A detailed timeline of the various data sets used in this manuscript can be found in Appendix F.

So far, we have discussed the behaviour at a fixed value of $V_{SG}$, which sets the electron density in the nanowire. We now turn our attention to the dependence of the ZBP on this gate parameter. Due to the cross-capacitance between the super and tunnel gates, changing $V_{SG}$ not only changes the density, but also the transmission of the junction. Although the reverse also takes place while changing $V_{TG}$, the influence of $V_{SG}$ affects the entire narrow tunnel junction while the cross-coupling effect of $V_{TG}$ is confined to only the superconducting segment immediately next to the junction. Therefore, to compensate for the more significant cross-coupling of $V_{SG}$, we simultaneously adjust $V_{TG}$ to keep $G_N$ approximately constant. In Figs. 6(a) and 6(b) we plot the conductance as a function of $V$ at 0.8 and 0.9 T, respectively, along the trajectory in $V_{SG}$-$V_{TG}$ space indicated by the dashed line in Fig. 6c. For both values of $B$, the ZBP persists over a sizable range in $V_{SG}$. At 0.8 T, the peak height first fluctuates around $2e^2/h$ with an amplitude of about 0.2×$2e^2/h$ (see also the line cuts in panel (d), left column). The decreasing conductance at $V$ = 0 for $V_{SG}$ > -5.1 V can mostly be attributed to the finite splitting of the ZBP. Similar splitting behaviour is observed at 0.9 T (see also the line cuts in panel (d), right column).

A similar scan at $B$ = 0.7 T is plotted in Fig. 7(a). A ZBP is present for most of the measured range, but with a height which significantly exceeds $2e^2/h$ (see also the line cuts in panel (d), left column). The number of charge jumps is also notably increased, causing $G_N$ to fluctuate and vary more strongly than in the previous figure. These instabilities make it difficult to get a consistent data set, which is demonstrated by the fact that the conductance at $V_{SG}$ = -6 V and $B$ = 0.7 T measured 16 hours later is lower by about 0.4×$2e^2/h$ (Fig. 7(b)).

Several detailed features can be observed inside the gap at finite energy, which appear to be interacting with the ZBP. In Fig. 7(c) we zoom in on the dashed rectangle in panel (a), showing an irregular behaviour of the ZBP with the peak height around $2e^2/h$ when the peak is at $V$ = 0 but dropping below this value when the peak is split (see also the line cuts in panel (d), right column). We speculate that this ZBP behaviour is due to several subgap states in the junction region[33].

We observe similar behaviour in device B, of which a scanning electron micrograph is shown in Fig. 8(a). This device is more stable, with less frequent charge jumps. The length of the proximitized section in this device is 0.9 μm, which is 0.3 μm shorter than in device A. Due to non-ideal device fabrication, the lower tunnel-gate was shorted to the left contact and thus kept at the potential of this lead during measurements. The upper tunnel-gate was shorted to the global back-gate and thus the same voltage was applied to both simultaneously. In Fig. 8(b) we plot the conductance as a function of $V$ and $B$, showing a ZBP with a maximum height slightly exceeding $2e^2/h$ at 0.83 T. In Figs. 8(c) and 8(d) we show the dependence on $V_{SG}$ and $V_{TG,BG}$, respectively, giving very similar results. The ZBP is present for a sizable gate range for both scans. In addition, both gates tune $G_N$ significantly, indicating a substantial capacitive coupling to the junction. The green dashed lines illustrate the region where $G_N$ is increasing while the ZBP height remains roughly constant, indicating a plateau-like behaviour. In this measurement, the peak height starts to increase further when the gate voltage is changed to more positive values. In a scenario with quasi-Majoranas, this increase could be due to an increased coupling to the second, more remote state[24]. Alternatively, in a scenario with MZMs this could indicate the opening of a second subband in the junction.

In Figs. 3, 4 and 8 we zoomed in on ZBPs taken at particular settings where the peak height has a tendency to stick to $2e^2/h$. We emphasize that, for different parameter settings, we can find ZBPs which have no tendency at all to stick to $2e^2/h$. As an example, we show a data set taken on device C in Fig. 9. The $B$ dependence (Fig. 9a) looks similar to the ones shown for devices A and B, with a ZBP-height close to $2e^2/h$ at 0.74 T. The ZBP persists over a sizable range in $V_{BG}$ (Fig. 9d). We also observe ZBPs for lower $B$-fields when $V_{BG}$ is changed (Fig. 9b). The ZBP-height increases monotonically in this case, and the value $2e^2/h$ does not appear to have a particular significance. The peak shape can be fitted with a Lorentzian line shape of varying height (Fig. 9f), with the peak height and width sharing a markedly different relation from the one shown in Fig. 4, see Appendix G. When $V_{TG}$ is changed, we also see that this peak is highly unstable, only appearing at $V = 0$ for very specific values (Fig. 9c). This type of level-crossing is similar to the typical Andreev bound state peaks reported in literature[34]. While we show these different types of behaviours in different devices, we expect that all of these behaviours could be measured in a single device if the parameter space is exhaustively searched. We note that the behaviour shown in Fig. 5 may be an intermediate case between a monotonically increasing peak height and the formation of a plateau-like peak height.

**Discussion**

Between the early reports on ZBPs in 2012, improvements in materials and fabrication[15,35-38] have yielded much cleaner tunnelling spectroscopy results. The induced gap has become 'hard' in the sense that no subgap states are visible at 0 T[32,35]. Andreev enhancement in QPCs has indicated interface transparencies exceeding 0.95[38]. In these cleaner devices, the ZBPs reach heights of order $2e^2/h$ relative to a low background conductance[30,39,40]. These observations strongly suggest that in clean devices tunnel spectroscopy can resolve individual quantum states. Moreover, unintentional (Coulomb blockaded) quantum dots can be excluded[15,38], which if hybridized with a superconductor, can also give rise to anomalies at zero bias[34].

Other forms of unintentional potentials, that could result in ZBPs, cannot easily be dismissed. One of us reported simulations assuming a random disorder potential (without Coulomb blockade) that resemble our experimental results[27]. Pan and Das Sarma dubbed this regime as the ugly one in the sequence "good, bad and ugly" ZBPs arising, respectively, from topological MZMs, quasi-MZMs, and disorder. Our experiments cannot exclude the ugly regime of random disorder. We do note that open questions remain. For instance, our ZBPs are very persistent in gate voltage, whereas random disorder may result in mesoscopic fluctuations in both peak height and splitting[26]. The scale in disorder, gate voltage, or chemical potential, for the occurrence of mesoscopic fluctuations is presently unknown, and whether disorder should induce mesoscopic fluctuations in these systems for the experimental parameter values is a relevant open question.

The 'bad regime' in Pan and Das Sarma is characterized by a smooth potential inducing quasi-Majorana (or ps-ABS) states in a magnetic field. In principle, quasi-Majoranas develop into topological MZMs when further increasing magnetic field. Signatures of such a transition, such as a closing and re-opening of the induced gap, have not yet been reported in the literature. Note that when the bulk Al gap closes before a topological transition occurs, then the quasi-Majoranas cannot transform to topological MZMs.

Numerical simulations show that quasi-Majoranas stick to zero energy over strikingly large ranges in the parameter space of ($B$, $V_{TG}$, $V_{SG}$)[23-25] . It is fairly easy to find parameter values for a smooth potential to simulate our experimental results with a high degree of resemblance, including the presence of plateau regions near $2e^2/h$. However, with fine tuning one can also generate plateau regions near $2e^2/h$ in the ugly regime[27]. This makes it hard to distinguish the ugly from the bad regime as long as we do not know our device parameters in sufficient detail. Tunnelling spectroscopy by itself thus appears to be unable to distinguish between disorder and quasi-MZMs as the origin of the observed ZBPs.

The "good regime" in Pan and Das Sarma contains a topological phase in the bulk and MZMs at the two opposite ends of the one-dimensional system. An experimental demonstration of this topological phase requires simultaneous spectroscopy on both the ends as well as the bulk. Preliminary results have been reported in a multi-probe device[41-43] but has the complication of possible unintentional potentials at each probe. A non-local scheme in a three-terminal geometry has been proposed[44] that would allow bulk and end spectroscopy for the same parameter settings.

In conclusion, we have reported large ZBPs with heights on the order of $2e^2/h$. Robust ZBPs sticking to zero energy has been reported earlier[9-15]. These studies, however, all reported limited peak heights, an order of magnitude less than $2e^2/h$. ZBPs with a height of order $2e^2/h$ were first reported from devices based on InAs 2DEGs[30]. Our large ZBPs are measured on InSb nanowires partially covered with Al. We frequently observe ZBPs with heights of order $2e^2/h$. In two devices we have found evidence for plateau behaviour where the peak height remains close to $2e^2/h$ while changing gate voltages over ranges where the normal conductance changes significantly. We note, however, that these plateaus are the result of a dedicated search motivated by proposals that such plateau behaviour would be indicative for MZMs[17,18,22]. Although, for a local spectroscopy experiment, we cannot think of another method, such a dedicated search has the potential to lead to confirmation bias and effectively yield false-positive evidence for MZMs. New numerical studies have shown that disorder-induced states and local, quasi-Majoranas can also yield plateau behaviour. Our data sets, nor any other data set reported in the literature that we know of, can make a distinction between the different possible origins dubbed as the "good, bad and ugly" in Ref. 27 since it would require knowledge of the microscopic potential landscape in the device. Our local measurements should be complemented with non-local conductance in order to be able to claim a correlation between a gap in the bulk of the nanowire while measuring ZBPs at the two ends[44].

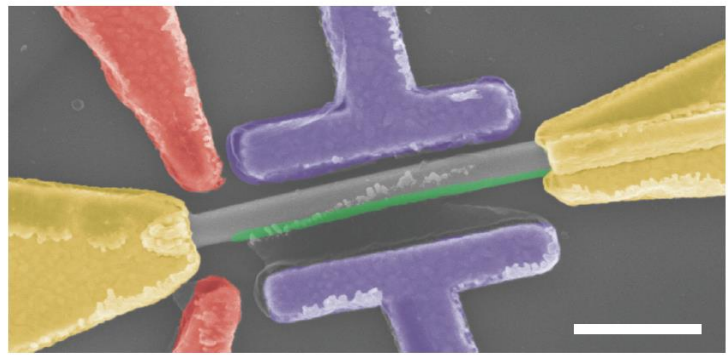

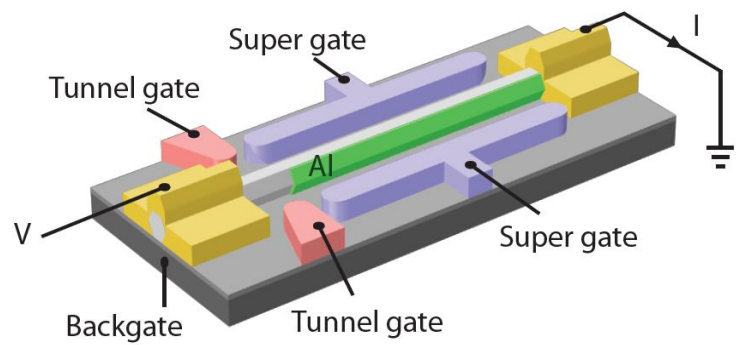

Figure 1. False color scanning electron micrograph of device A (upper panel) and its schematics (lower panel). An InSb nanowire (grey) is covered on two of its six facets by a thin superconducting Al shell (green) with a thickness of approximately 10 nm. Side gates and contacts are Cr/Au (10 nm/100 nm). The substrate is p-doped Si, which can be used as a global backgate, covered by 285 nm $SiO_2$. The two tunnel gates (red) are shorted externally, as are the two super gates (purple). A bias voltage $V$ is applied to the contact (yellow) on the left and a current $I$ is drained from the contact on the right. Scale bar is 500 nm.

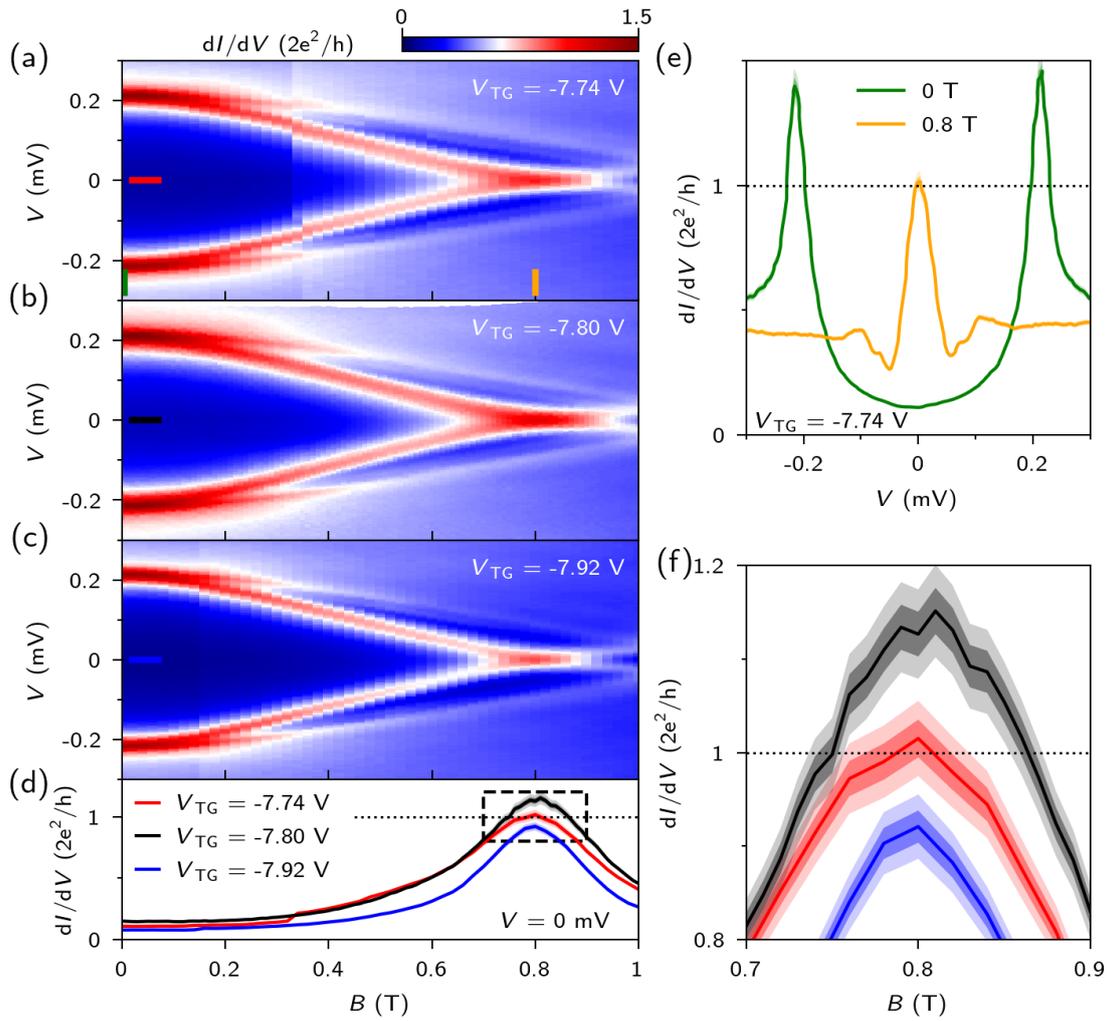

Figure 2. (a-c) d$I$/d$V$ versus $V$ and $B$ at three different settings of $V_{TG}$. $B$ is aligned with the nanowire axis for all measurements. $V_{SG}$ = -6.5 V, $V_{BG}$ = 0. Fridge base temperature is ~20 mK for all measurements unless specified. (d) Horizontal linecuts at $V$ = 0 mV for panels a-c. (e) Vertical linecuts from panel a, at $B$ = 0 T (green) and 0.8 T (orange). (f) Close-up of the data in panel (d), indicated by the dashed rectangle. Darker shading indicates the 1σ standard error in the conductance measurement, while the lighter shade indicates the 2σ error. Data was obtained from device A.

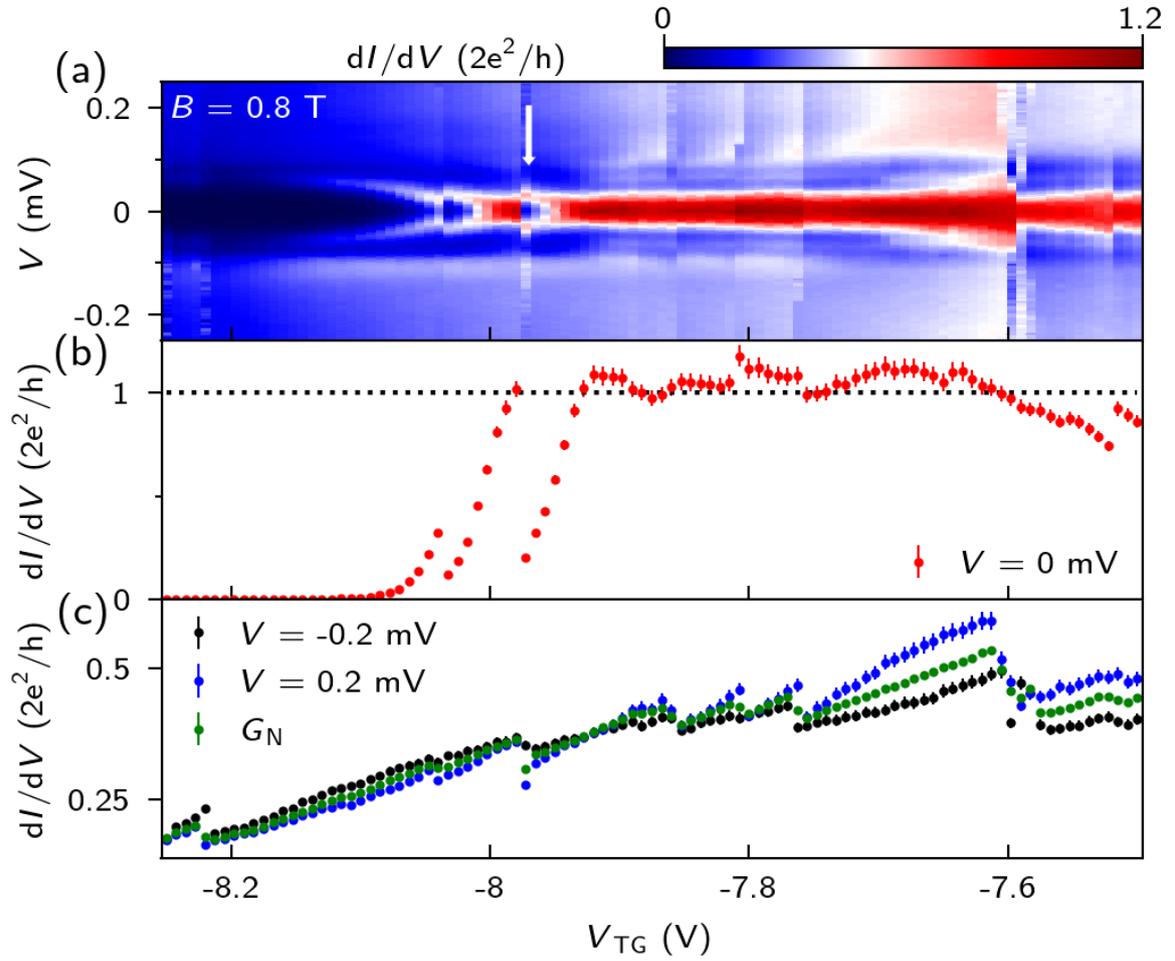

Figure 3. (a) d$I$/d$V$ versus $V_{TG}$ at $B$ = 0.8 T, $V_{SG}$ = -6.5 V, $V_{BG}$ = 0. White arrow indicates a charge jump with a particularly large effect on the conductance. (b) Horizontal linecut from (a) at $V$ = 0 mV, showing the height of the ZBP remains roughly constant over a range in $V_{TG}$. (c) Horizontal linecuts at finite bias, ±0.2 mV (black, blue), showing the out-of-gap conductance increasing over the same range in gate voltage. We further characterize the out-of-gap conductance by taking the average value for $|V| \geq 0.2$ mV, which we label $G_N$ (green). Error bars are given at 2σ for clarity. Data was obtained from device A.

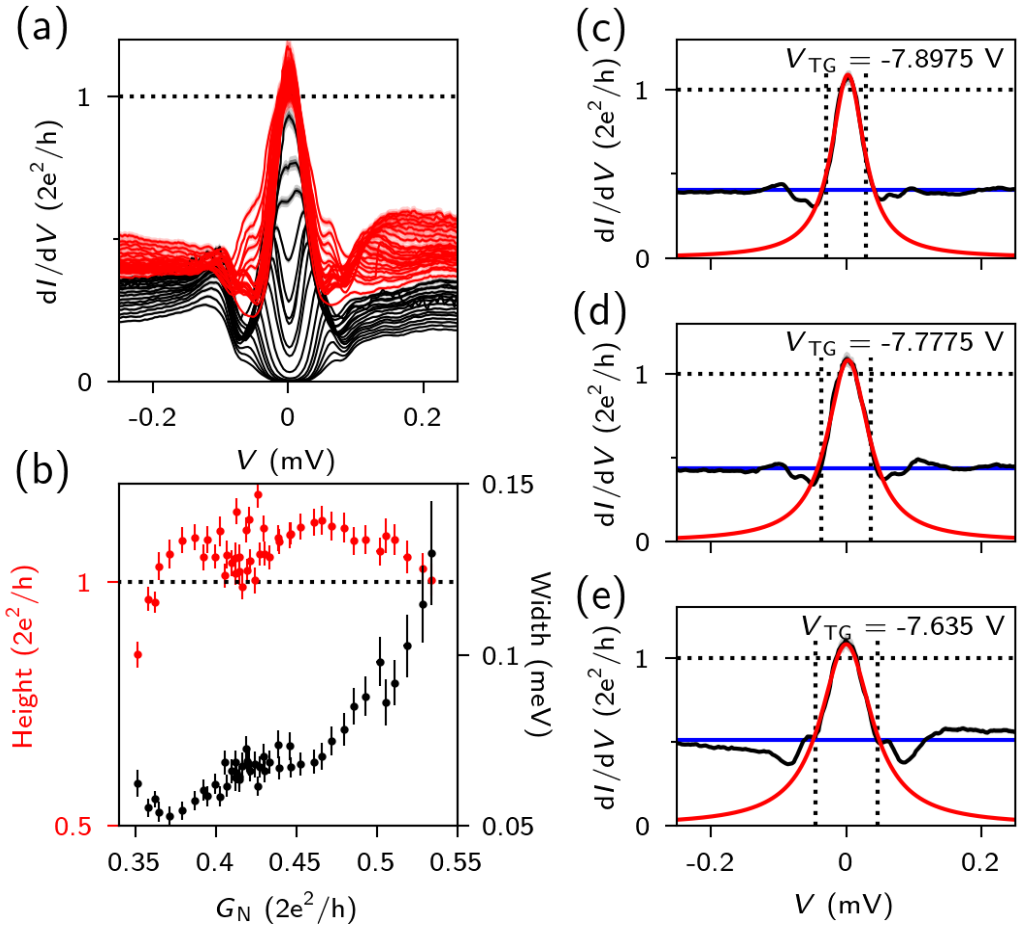

Figure 4. (a) d$I$/d$V$ versus V traces corresponding to linecuts in Fig. 3(a), showing the ZBP curves for $V_{TG}$ ranging from -7.9275 V to -7.6125 V (red curves, ZBP height > $0.95 \times 2e^2/h$). Due to peak splitting, the ZBP height drops lower for more negative $V_{TG}$ (black curves). To avoid crowding, every other linecut is plotted. (b) ZBP height (red circles) and ZBP width (black circles) extracted from all linecuts of Fig. 3(a) for -7.9275 < $V_{TG}$ < -7.61 V, as a function of above-gap conductance ($G_N$). The ZBP height and width are determined by fitting a Lorentzian shape to the data, several examples of which are shown in panels (c-e) (red lines). The blue lines indicate the extracted value of $G_N$ for the same gate voltage. Error bars in panel (b) are given at 2σ.

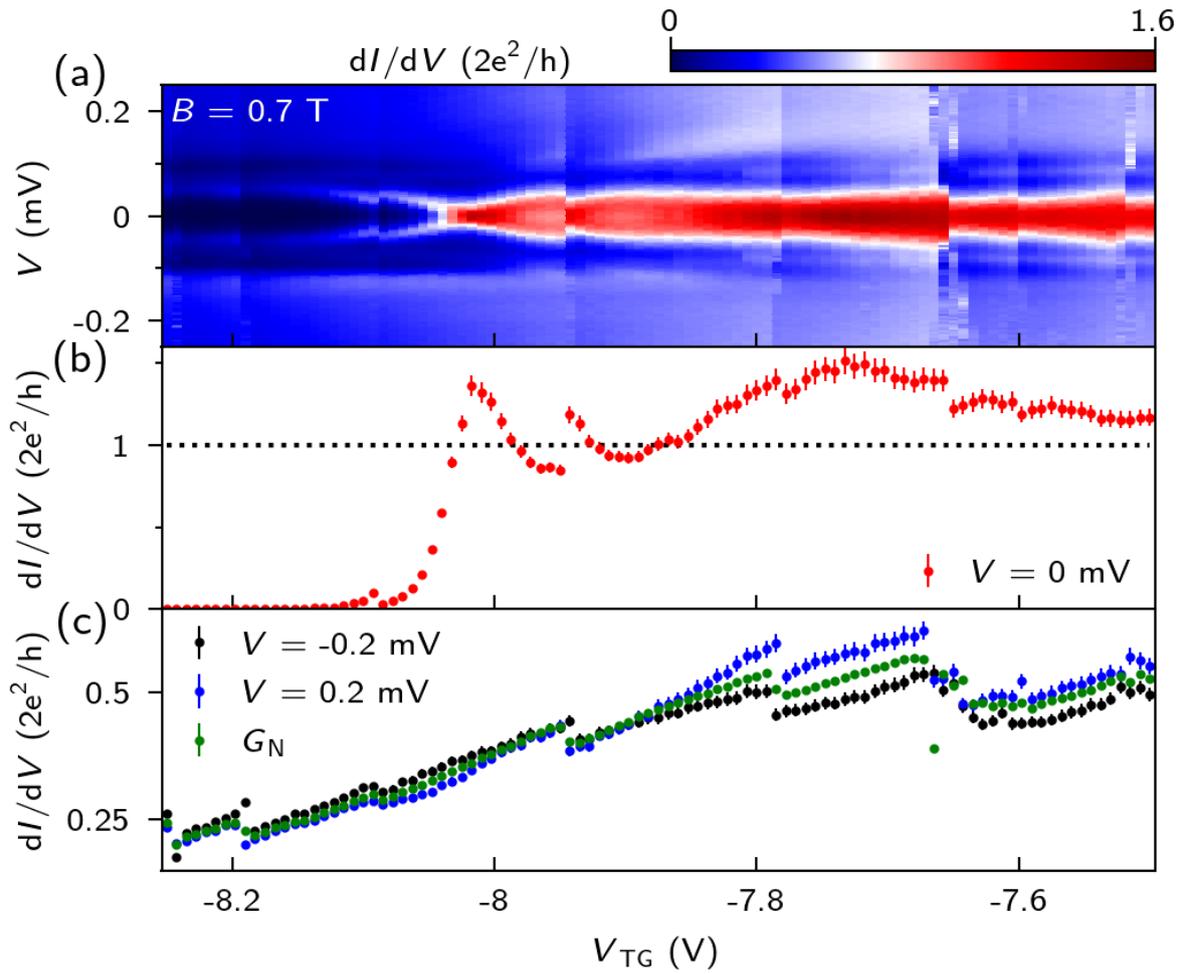

Figure 5. (a) d$I$/d$V$ versus $V_{TG}$ at $B$ = 0.7 T, $V_{SG}$ = -6.5 V, $V_{BG}$ = 0. (b-c) Horizontal linecuts from (a), showing ZBP height and above-gap conductance, respectively. The ZBP height exceeds $2e^2/h$ considerably at this field value. Data was obtained from device A.

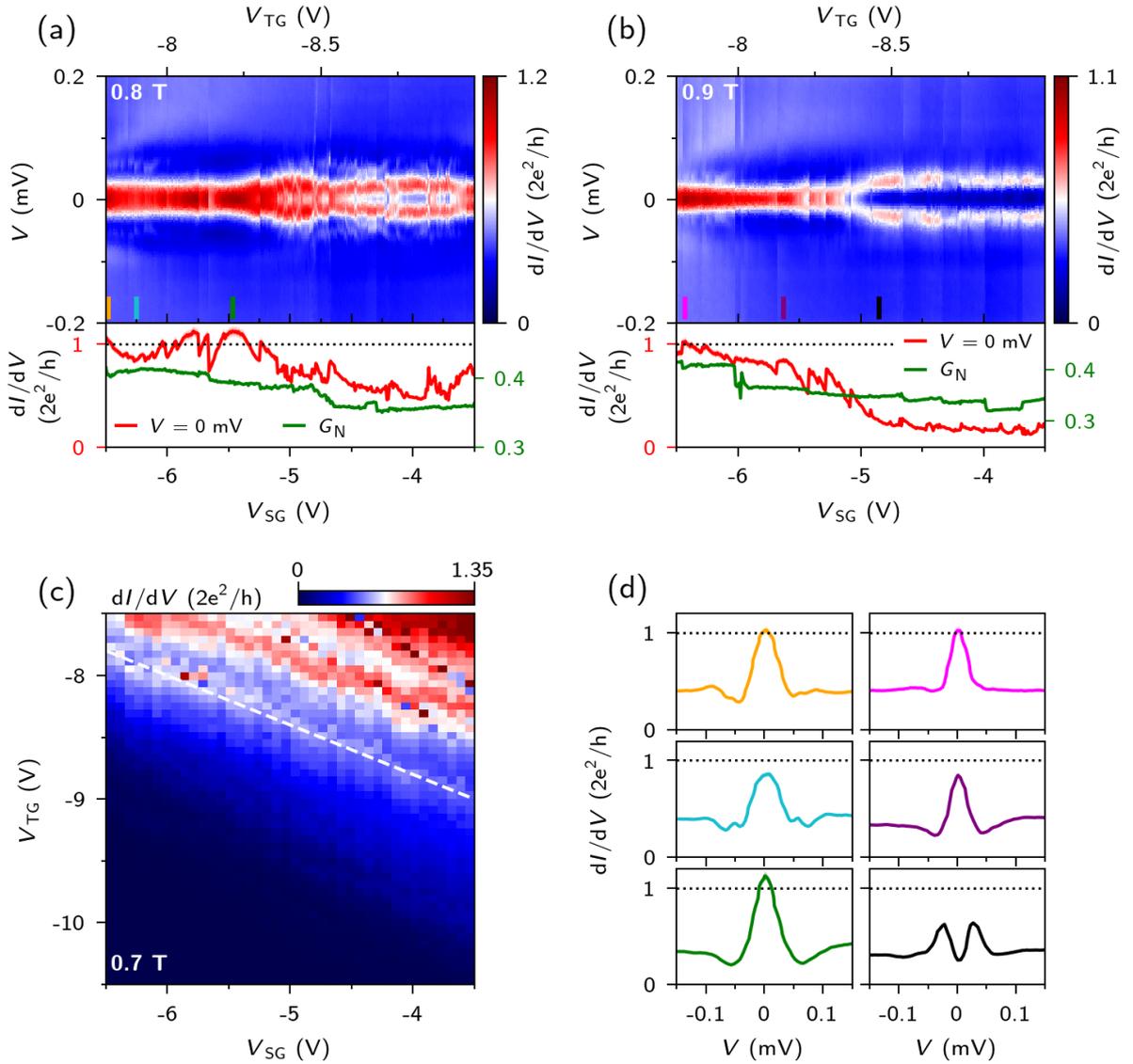

Figure 6. (a-b) d$I$/d$V$ versus $V_{SG}$ at $B$ = 0.8 T and 0.9 T, respectively. Lower panels show the ZBP height (red curve) and $G_N$ (average conductance for $|V| \geq 0.2$ mV, green curve), with the (almost invisible) shaded region representing the error bar. $V_{TG}$ is adjusted while sweeping $V_{SG}$, as shown on the top axis, to keep the above-gap conductance roughly constant. (c) Conductance at $V$ = -0.5 mV as a function of $V_{TG}$ and $V_{SG}$ at 0.7 T. The gate voltage scans in panels a and b follow the white dashed line. (d) Several vertical linecuts from panes a and b showing ZBPs and split peaks at gate voltages and magnetic fields indicated by the correspondingly colored bars. Data was obtained from device A.

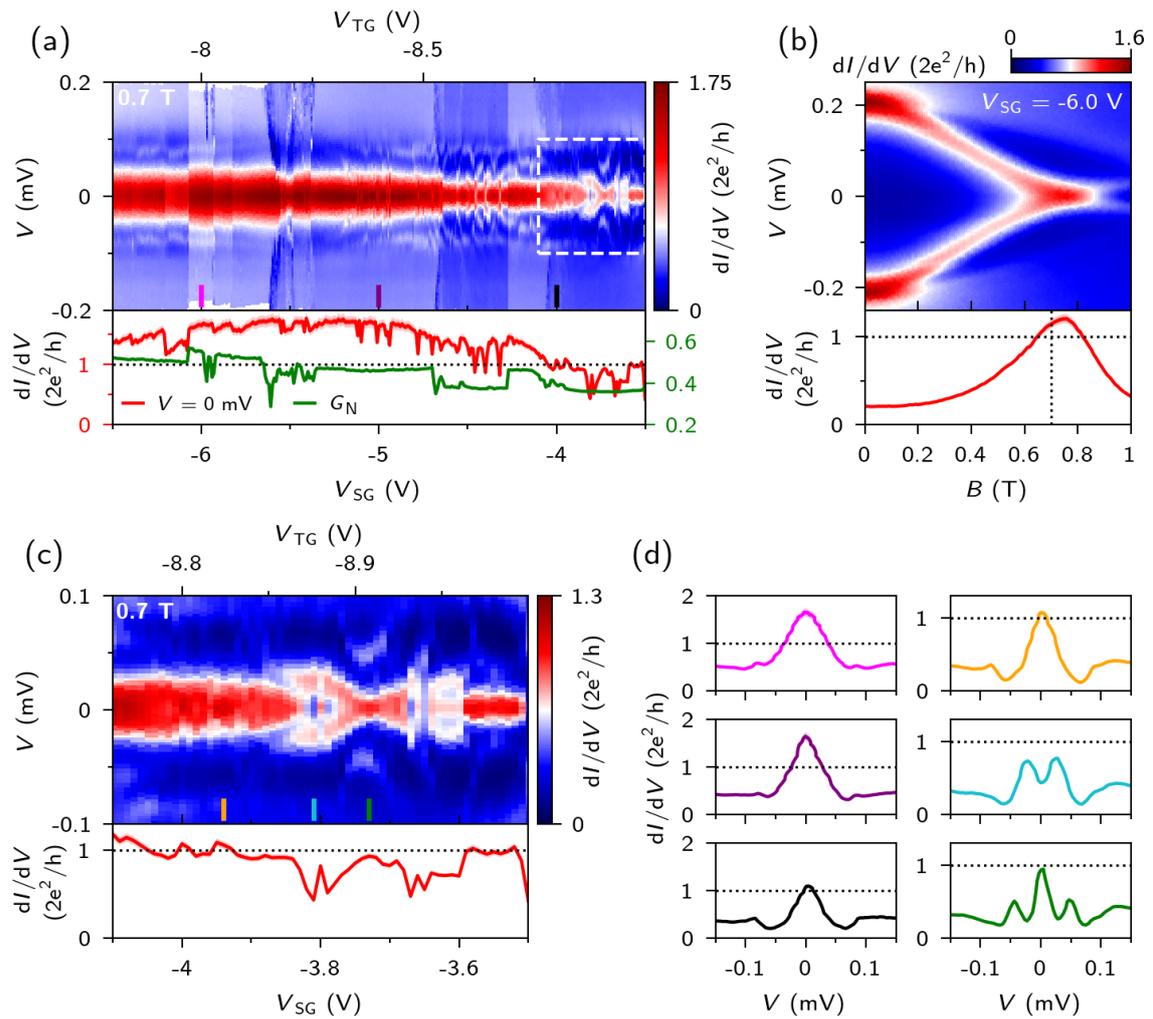

Figure 7. (a) d$I$/d$V$ versus $V_{SG}$ at $B$ = 0.7 T, with $V_{TG}$ compensation following the white dashed line in Fig. 6c. Lower panel shows the ZBP height (red curve) and $G_N$ (average conductance for |$V$| ≥ 0.18 mV, green curve), respectively. The ZBP height significantly exceeds $2e^2/h$, consistent with the peak height measured in Fig. 5. (b) Top panel shows $B$ dependence of a ZBP whose height significantly exceeds $2e^2/h$, at $V_{SG}$ = -6.0 V, $V_{TG}$ = -8.0 V. Bottom panel shows horizontal linecuts at $V$ = 0 mV, showing the $B$ dependence of ZBP height. (c) Zoom-in of the white dashed rectangle in panel a, where the peak height is close to $2e^2/h$. The ZBP in this gate range shows oscillatory (peak vs split peak) behavior. (d) Several vertical linecuts from a-c at gate voltages indicated by the correspondingly-colored bars. Data was obtained from device A.

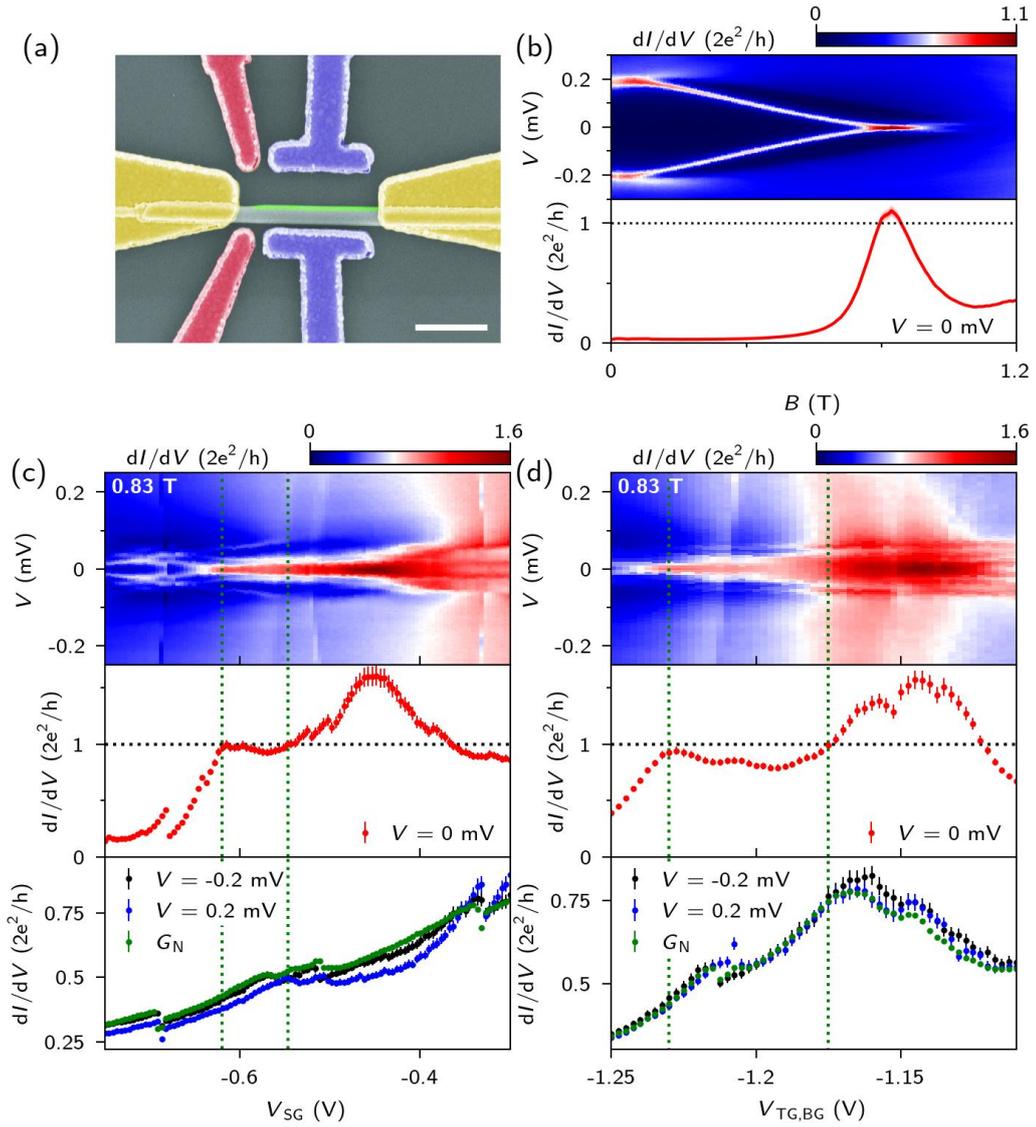

Figure 8. (a) False color scanning electron micrograph of device B. In this device, the lower tunnel gate is shorted to the left lead and thus follows its potential during the measurement. The upper tunnel gate is shorted to the backgate, and thus follows the potential of the backgate, $V_{TG} = V_{BG}$. Scale bar is 500 nm. (b) d$I$/d$V$ vs $V$ and $B$. Bottom panel shows the horizontal linecut at $V = 0$ mV, showing behavior similar to device A with the maximum of the peak height slightly exceeding $2e^2/h$. (c) d$I$/d$V$ versus $V_{SG}$ at 0.83 T. The two lower panels are the horizontal linecuts at $V = 0$ mV (red line), and ±0.2 mV (blue and black lines), as well as the average normal state conductance $G_N$. The two dashed green lines indicate a plateau region of the zero-bias conductance. $V_{TG,BG} = -1.23$ V. (d) d$I$/d$V$ vs $V_{TG,BG}$ at 0.83 T. The two lower panels show the zero-bias and normal state conductance, similar to panel c. The two dashed green lines indicate the plateau region of the zero-bias conductance. $V_{SG} = -0.615$ V. Error bars are given at the 2σ level. Data was obtained from device B.

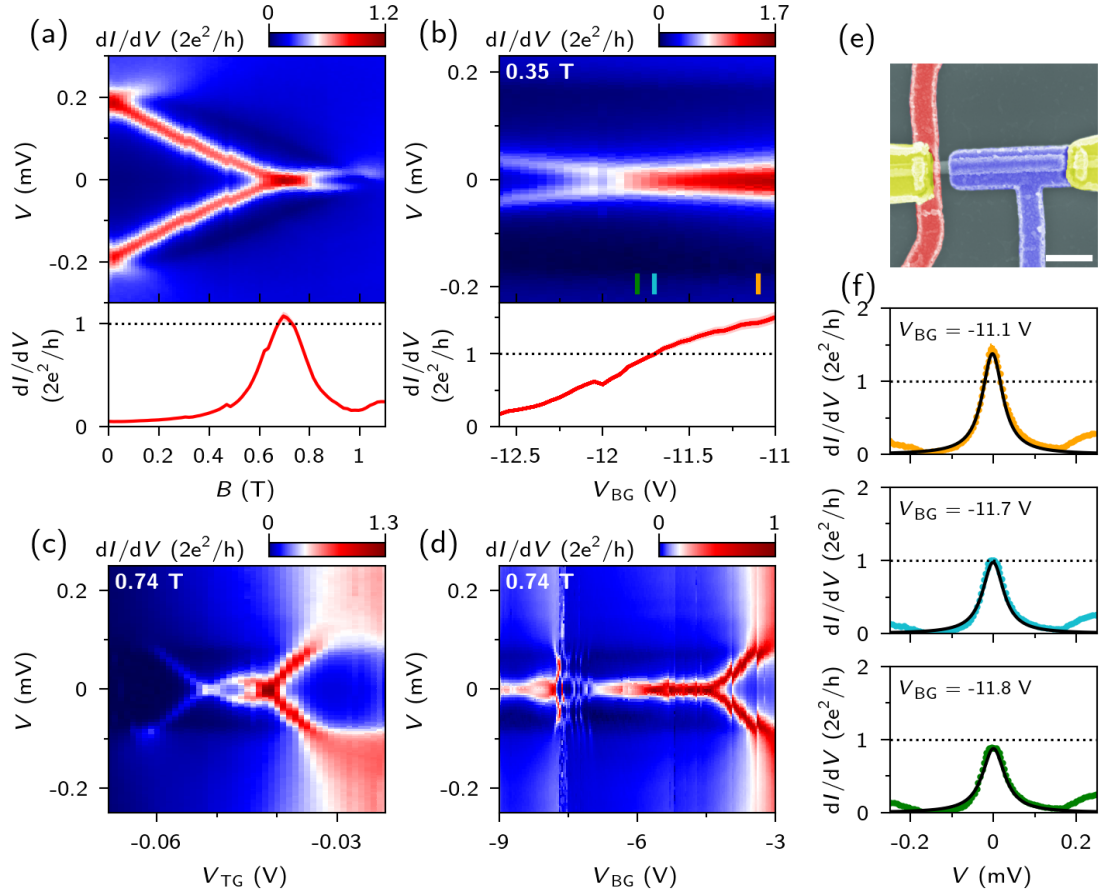

Figure 9. (a) d$I$/d$V$ versus $V$ and $B$. Bottom panel shows the horizontal linecut at $V$ = 0 mV. $V_{BG}$ = -5.58 V, $V_{TG}$ = -0.045 V, $V_{SG}$ = 0.045 V. (b) d$I$/d$V$ versus $V$ and $V_{BG}$ (with more negative gate voltage values compared to (a)) at a lower field of 0.35 T. $V_{TG}$ = 0.01 V, $V_{SG}$ = 0 V. The lower panel shows a zero-bias linecut where the ZBP height monotonically increases and crosses $2e^2/h$ without any particular feature. (c) d$I$/d$V$ versus $V$ and $V_{TG}$ at 0.74 T, showing the peak as a result of a level crossing. $V_{BG}$ = -5.58 V, $V_{SG}$ = 0 V. (d) d$I$/d$V$ versus $V$ and $V_{BG}$ at 0.74 T, showing the peak remains unsplit over a sizable gate voltage range with the presence of several charge jumps. $V_{TG}$ = -0.03 V, $V_{SG}$ = 0 V. (e) False color scanning electron micrograph of device C. The top gates, separated from the nanowire by 30 nm thick SiN dielectric, serve as the tunnel gate (red) and super gate (purple). Scale bar is 500 nm. (f) Vertical linecuts from panel b, at the gate voltages indicated by the colored lines. Lorentzian fits are shown in black. Error bars are given at the 2σ level. Data was obtained from device C.

# Large zero-bias peaks in InSb-Al hybrid semiconductor-superconductor nanowire devices

Hao Zhang*, Michiel W.A. de Moor*, Jouri D.S. Bommer*, Di Xu, Guanzhong Wang, Nick van Loo, Chun-Xiao Liu, Sasa Gazibegovic, John A. Logan, Diana Car, Roy L. M. Op het Veld, Petrus J. van Veldhoven, Sebastian Koelling, Marcel A. Verheijen, Mihir Pendharkar, Daniel J. Pennachio, Borzoyeh Shojaei, Joon Sue Lee, Chris J. Palmstrøm, Erik P.A.M. Bakkers, S. Das Sarma, Leo P. Kouwenhoven[†]

# Appendix

## Contents

# Appendix A. Additional devices

Over 60 devices were fabricated and tested over the course of the project. Out of these, we selected 11 devices, which showed good transport characteristics and had functional electrostatic gates, for extensive measurements. Although most of these devices show ZBPs after fine tuning gate voltages and magnetic field, some even with peak heights close to $2e^2/h$, only 2 devices (Device A and B) show behavior consistent with a plateau in the ZBP height in a specific parameter range. The other devices show behavior similar to what is shown in Fig. 9 (data taken from device C). Scanning electron micrographs of devices A, B and C are shown in Fig. 1, Fig. 8a and Fig. 9e, respectively. In Fig. A1 we show the scanning electron micrographs and device schematics of the other 8 devices.

Devices 1 and 2 are side-gated devices, similar in design to devices A and B. Device 3 is covered by 30 nm of SiN dielectric and has a local tunnel gate as well as a global backgate separated from the nanowire by 20 nm thick LPCVD $Si_3N_4$. For all other devices, the global backgate is covered by 285 nm of $SiO_2$. Devices 4 and 5 have a similar design to device C, with a tunnel gate and super gate on top, separated from the nanowire by 30 nm of SiN dielectric. Devices 6 to 8 have two layers of gates and dielectric. The nanowire is first covered by 30 nm of SiN and the local tunnel gate, which are subsequently covered by another layer of 30 nm SiN and the super gate.

Since this work was first conducted, several works have been published investigating the role of device geometry and gate design on the topological phase diagram through Schrödinger-Poisson simulations, see e.g. ref [1, 2]. The device geometry can also play a role in generating Andreev bound states due to potential inhomogeneities, see e.g. ref [3, 4].

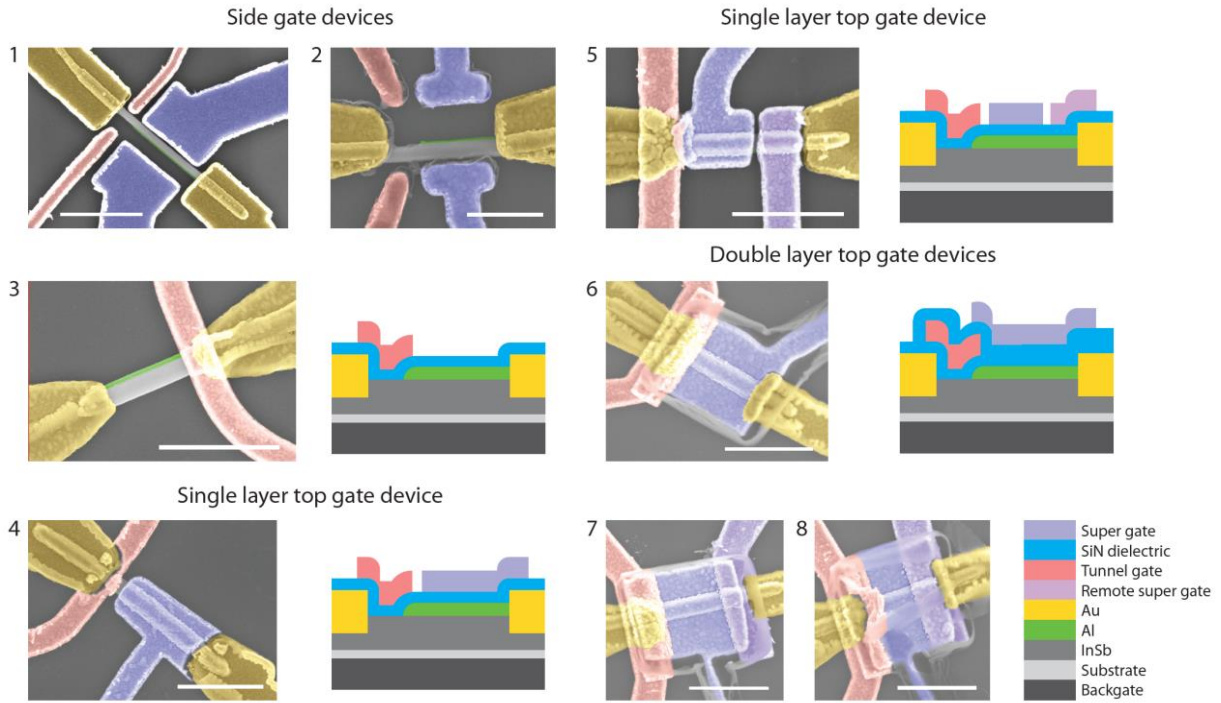

Figure A1. Scanning electron micrographs and design schematics of 8 additional devices studied during this project. The scale bar is 1 µm for all panels except for device 2, where it is 500 nm.

# Appendix B. Measurement circuit and conductance calculation

The experiments described in this work consist of two-terminal voltage-biased conductance measurements. A schematic representation of the measurement circuit is shown in Fig. B1. The device is mounted onto a PCB with integrated RC filters (orange dashed box). This PCB is mounted to the mixing chamber plate of a dilution refrigerator, with additional RC filters (blue dashed box). In addition to the RC filters, pi filters and copper powder filters are installed to filter high frequency noise. We apply both a DC and AC voltage to the sample through an in-house built S3b voltage amplifier [5]. The resulting current is amplified by an in-house built M1b current-to-voltage converter [6]. The DC component of the voltage output of the current-to-voltage converter is measured with a Keithley 2000 multimeter, while the AC component, consisting of an in-phase (X) and quadrature (Y) signal, is detected using an SR830 lock-in amplifier.

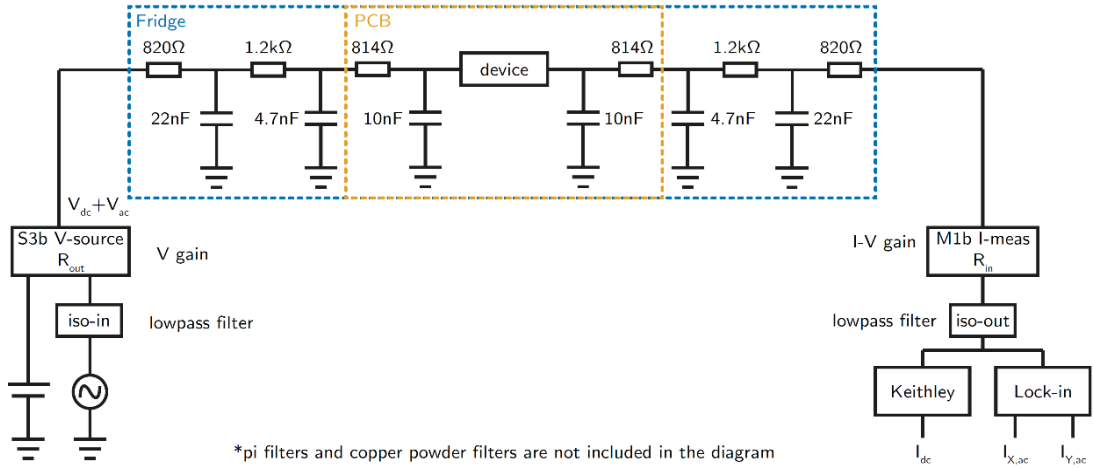

Figure B1. Schematic representation of the measurement circuit. Note that each fridge line includes additional pi filters and copper powder filters which are not shown (cutoff frequencies >10 MHz and negligible resistance).

Because of the additional series resistances in the circuit, the actual voltage across the device, $V$, is less than the applied voltage $V_{dc}$:

$$V = V_{dc} - I_{dc} R_{series}$$

where $R_{series}$ is the sum of the filter resistances in the circuit, as well as the input impedance of the current-to-voltage converter and the output impedance of the voltage source. We have independently calibrated this series resistance. This resistance also needs to be accounted for when calculating the differential conductance, $dI/dV$, giving

$$\frac{dI}{dV} = \frac{I_{ac}}{V_{ac} - I_{ac} R_{series}}$$

As we are interested in the response in the dc limit, ideally one would take only the in-phase component of the ac current, $I_{ac} = I_{X,ac}$, to calculate the differential conductance. However, due to the limited bandwidth of the current-to-voltage converter, a phase shift is introduced which means the quadrature component $I_{Y,ac}$ cannot be neglected. Additionally, the gain of the converter is reduced, resulting in an underestimation of the total AC current. Due to additional reactive elements in the circuit, such as the RC-filters, the amount by which the current is underestimated also depends on the device conductance, making it very complicated to extract this quantity directly from the measured signals. To remedy these effects, we have used a bootstrapping method to calibrate the measured ac conductance to the numerical derivative of the dc signal. This method is explained in detail in ref. [7] with access via the online data repository [8]. For additional information regarding the measurement of differential conductance using a lock-in amplifier, see ref. [9]. The uncertainties introduced by this procedure, as well as the uncertainties in the independently measured values of the series resistance in the circuit and the gain of the current-to-voltage converter, have been included in the error bars for the presented conductance data.

## Appendix C. Discussion on contact resistance

As discussed in Appendix B, because we conduct our experiments in the two-terminal configuration, we need to account for additional series resistances in the circuit when calculating the conductance through our devices. The resistances in the measurement circuit are known, but it is possible that there is an additional barrier at the interface between the metal leads and the semiconductor nanowire, generally called "contact resistance". Depending on the materials and fabrication procedures used, this interface can go from a highly-resistive Schottky-type contact, to a moderately transparent Ohmic contact, to fully transparent. As long as the applied bias voltage is small, it is generally assumed that the contact resistance is independent of the voltage, i.e., Ohmic. While ideally one would use a four terminal configuration to circumvent this problem, the nanowires used in this work are too short to allow for four independent contacts, which means that at best a quasi-four terminal configuration can be used where a single contact on the wire is split into two separate lines on the PCB. While this eliminates the effect of the known series resistances in the circuit, it leaves the contact resistance undetermined.

A typical method to estimate the contact resistance is by matching the conductance plateaus in a quantum point contact (QPC) to their quantized values. However, this requires at least two plateaus to be present. Alternatively, in an Andreev QPC the role of the second plateau can be taken by the enhanced conductance at $V = 0$. For a single channel Andreev QPC with a transmission $T$, these two quantities are related by the Beenakker formula [10] (assuming $\mu \gg \Delta$ [11])

$$G(V = 0) = G_S = \frac{2e^2}{h} \frac{2T^2}{(2-T)^2}$$

As $T$ can be deduced from the value of the conductance plateau outside the superconducting gap, in principle one could use this formula to determine the series resistance as a fitting parameter. As an alternative approach, one can use the DC signal and the size of the superconducting gap as an independent estimate. The measured superconducting gap is not expected to change with $V_{TG}$. If, however, the wrong series resistance is subtracted, the superconducting gap will appear to change with $V_{TG}$ as the bias axis becomes distorted.

In Fig. C1 we present an Andreev QPC measured in device C. In panel a, we show the extracted differential conductance before any series resistances in the circuit are accounted for. The edge of the superconducting gap appears to move to higher bias as the gate voltage is increased, because the effective voltage drop over the junction decreases when the device conductance increases. When we correct for the independently calibrated series resistance (for this measurement, 17.152±0.089 kΩ), this effect disappears. In Fig. C1c, we see that for low transmission the system shows a hard gap (bottom), while on the plateau the conductance is strongly enhanced inside the gap (top). Using the BTK model [12], we can fit the conductance as a function of $V$ to extract the superconducting gap, as well as the transmission $T$, which is related to the BTK $Z$ parameter via $T = 1/(1+Z)^2$. We plot the extracted transmission in Fig. C1d (red dashed line), in addition to the out-of-gap conductance $G_N$ (green) and the conductance at $V = 0$, $G_S$ (purple). The expected conductance $G_S$ based on the extracted transmission and the Beenakker formula is plotted as the blue dashed line. In Fig. C1d we plot the values of $G_S$ as a function of the transmission (black dots), showing excellent agreement with the Beenakker formula (red line), demonstrating that this data set can be described accurately by a single ballistic channel connecting a normal and a superconducting reservoir without subtracting a contact resistance. As the transmission on the plateau already closely approaches the value of 1, it is unlikely that a significant additional series resistance remains unaccounted for in the circuit.

Alternatively, using the methods of minimizing the variation of the superconducting gap with $V_{TG}$, as well as matching the plateau values of the Andreev QPC, we find that an optimal series resistance between 17 and 17.7 kΩ, depending on the specifics of the fitting procedure. Based on these results, as well as results obtained from other devices, we consider it likely that the total contact resistance is in the range of a few hundred Ohms, possibly up to 1 kΩ. We have not corrected for this, as the described procedures contain several assumptions (e.g. whether or not the contact barrier is Ohmic). It should be noted that this means the reported conductance values could be underestimated.

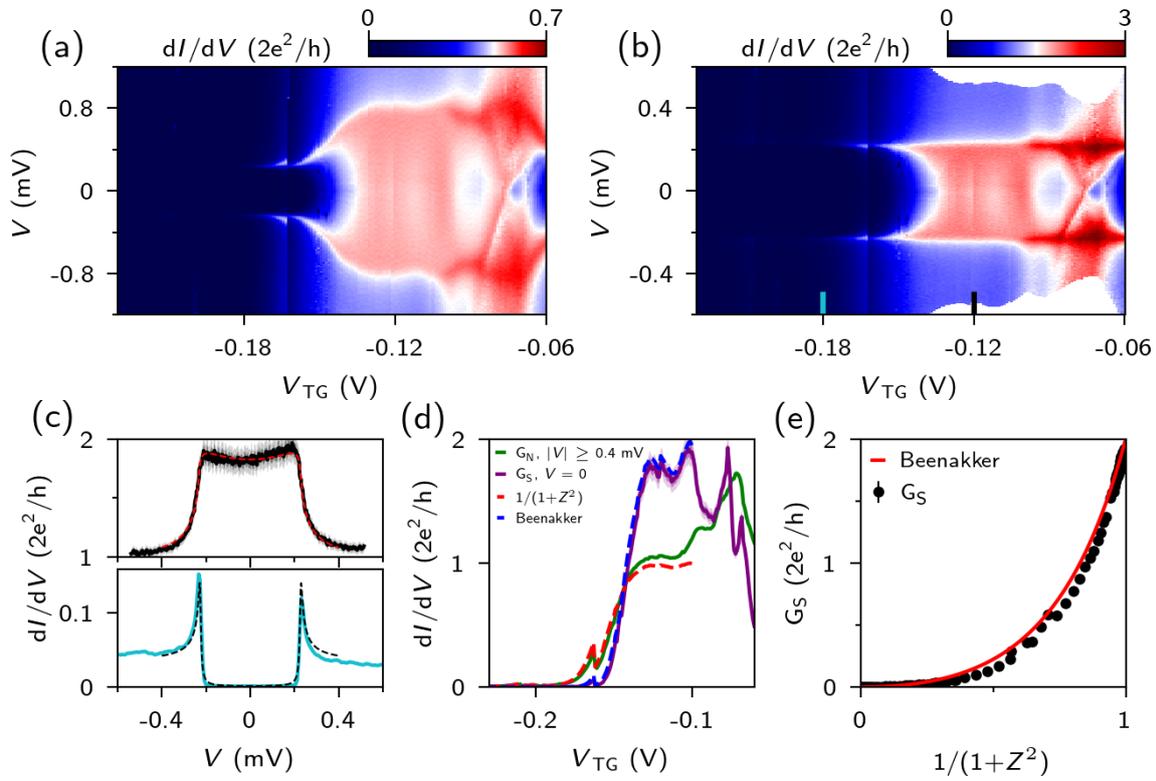

Figure C1. Ballistic Andreev transport. (a) d$I$/d$V$ versus $V$ and $V_{TG}$ before correcting for a series resistance in the circuit. (b) d$I$/d$V$ versus $V$ and $V_{TG}$ after correcting for the known series resistances. (c) Line trace at $V_{TG}$ = -0.18 V (bottom) and $V_{TG}$ = -0.12 V (top), showing a hard gap and enhanced conductance due to Andreev transport, respectively. Dashed lines are fits to the BTK model. (d) Out-of-gap conductance $G_N$ (averaged for $|V| \geq 0.4$ mV, green) and conductance at $V$ = 0, $G_S$. Dashed lines indicate the transmission $1/(1+Z)^2$ as extracted from the BTK fit (red), and the predicted zero-bias conductance from the Beenakker formula (blue). (e) $G_S$ plotted versus the transmission $1/(1+Z)^2$ (black dots), showing excellent agreement with the Beenakker prediction (red line). Data was obtained from device C.

## Appendix D. Discussion on charge jumps

In addition to designed elements such as the metal gates, the electrostatic environment of a mesoscopic device can be affected by the presence of local charge traps. These traps can be located e.g. in the dielectric materials or in residues left over from the fabrication process. As gate voltages are changed during the experiment, these traps can be charged or discharged. When there are many traps present, this will generally result in changing electric fields which are visible in the experiment as effective gate drifts and hysteresis. When a single trap suddenly charges or discharges itself in such a way that the effective gate potential is suddenly shifted, we speak of a "charge jump". These charge jumps can occur stochastically, or in a reproducible fashion where the charging and discharging process is triggered by the application of a specific gate voltage. Such charge rearrangements are well-known properties of semiconductor electronic devices in general, not just the specific devices studied in this work.

In Fig. D1, we show an example of a non-reproducible, stochastic charge jump. In panel a, the gate voltage is slowly decreased until $V_{TG}$ = -0.163 V, indicated by the white arrow. At this point, a charge jump occurs, which shifts the gate potential by about 0.007 V. When the same measurement is conducted again, the charge jump is no longer present, as shown in Fig. D1b. In this panel, we see other features also seen in panel a.

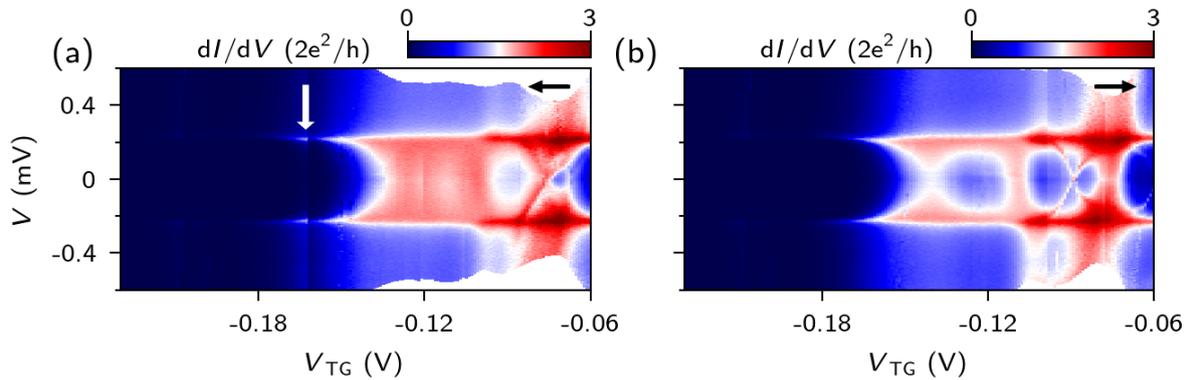

Figure D1. (a) Same as Fig. C1b. The white arrow indicates a charge jump which causes a shift in $V_{TG}$. (b) The same measurement taken consecutively, where the charge jump is now absent. Black arrows indicate the direction in which the gate voltage is changed over the course of the measurement. Data was obtained from device C.

Another difference is the change in Andreev enhancement on the plateau, which is very sensitive to precise details of the gate potential and therefore the most likely feature to be affected by a jump. In general, a charge jump will not exactly correspond to a change in the gate potential, but instead constitute a more complex (and unknown) shift in the local electric field. However, when the measured conductance is almost exactly repeated for every bias voltage, it stands to reason that the particular charge jump in question approximates a shift in the applied gate voltage. We illustrate this situation in Fig. D2a. In black, we plot the seven traces at fixed $V_{TG}$ to the left of the charge jump, while in red we plot the seven traces to the right. Every trace to the left of the charge jump is almost identical to one shifted by 0.007 V to the right. We can see this more clearly in Fig. D2b, where we have plotted the conductance as a function of $V_{TG}$ for several values of the bias voltage $V$ (black lines). If we shift these traces by 0.007 V (dashed red lines), we see almost an exact match in the overlapping region. This indicates that this particular charge jump indeed approximates a shift in the applied gate voltage.

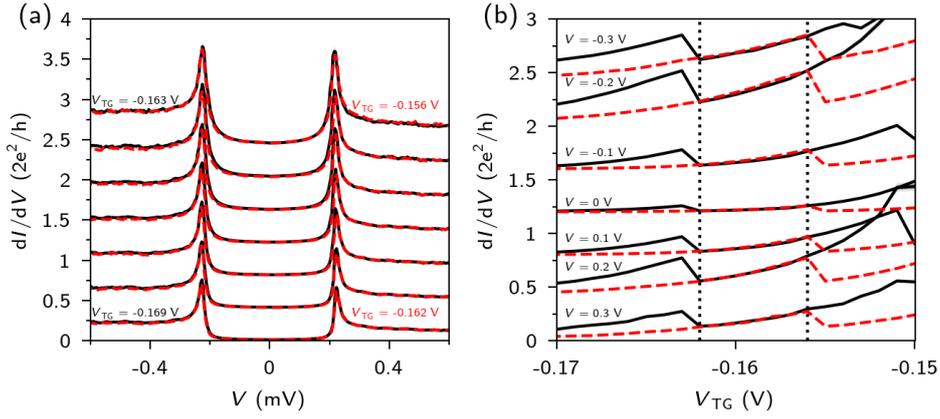

Figure D2. (a) Line traces at fixed $V_{TG}$ from Fig. D1a, showing the data just to the left of the charge jump (black) and to the right (red). Traces are offset by 0.8 $e^2/h$ for clarity. (b) Line traces at fixed $V$ from Fig. D1a (black), and the same line traces shifted by 0.007 V (red). In the overlapping region (indicated by the vertical dotted lines), the two traces match for every bias voltage.

A similar analysis can be used to show that the charge jump indicated by the white arrow in Fig. 3 corresponds to an effective shift in the tunnel gate voltage, shown in Fig. D3. In contrast, the jumps in the conductance which can be observed at $V_{TG}$ = -7.86, -7.81, and -7.76 V do not cause the conductance values to repeat, suggesting that these jumps do not correspond to effective shifts in the tunnel gate potential. They could instead be caused by single charged impurities or resonances near the NS-junction. The fact that similar jumps are observed in Fig. 5, albeit at slightly different gate voltages, suggests that these are reproducible resonances in a certain part of the parameter space, the energy levels of which can be altered by magnetic field or gate voltage sweeps.

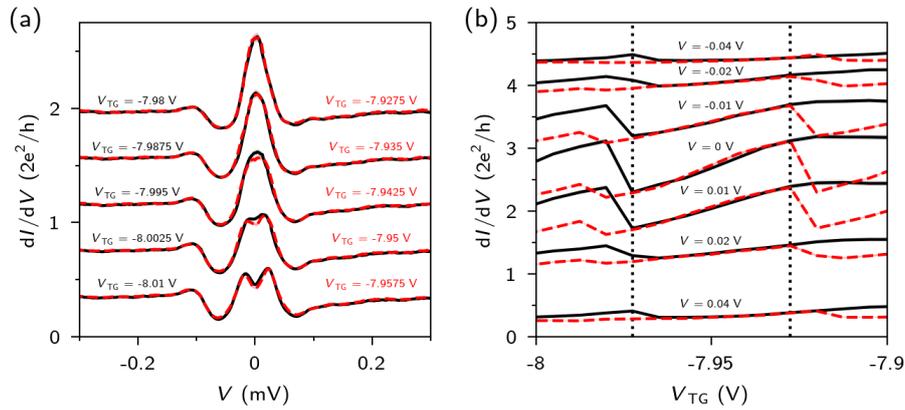

Figure D3. (a) Line traces at fixed $V_{TG}$ from Fig. 3a, showing the data just to the left of the charge jump marked by the white arrow (black) and to the right (red). Traces are offset by 0.8 $e^2/h$ for clarity. (b) Line traces at fixed $V$ from Fig. D2a (black), and the same line traces shifted by 0.0525 V (red). In the overlapping region (indicated by the vertical dotted lines), the two traces match for every bias voltage.

# Appendix E. Discussion on plateau-like features

Previous studies of zero-bias peaks have mainly focused on the presence of a peak at zero bias voltage. The next step is to consider the height of the peak. If the height of a ZBP remains constant while changing parameters, it is referred to as a "plateau". If the plateau value is close to $2e^2/h$, it is called a "quantized plateau". We emphasize that both a robust ZBP and a stable quantization are signatures of MZMs although not sufficient for an unambiguous proof of their existence.

A plateau is generally considered to be a region of parameter space in which the value of some quantity remains constant to within some acceptable tolerance. In the context of Majorana zero modes, the constant quantity is the height of the zero-bias conductance peak, with the region of parameter space being a subset of the parameter space of the topological phase, with the exact extent depending on additional details such as the relevance of finite size or temperature. Additionally, in the case of a sufficiently long, ideal 1D single subband Majorana wire and for temperatures smaller than the lifetime broadening (i.e. FWHM of ZBP) [13], the value of the conductance plateau should be quantized to the universal value of $2e^2/h$ to within experimental accuracy.

The total parameter space in our experiments is far too large to search exhaustively. It is clear that given the number of experimental tuning parameters, some well-defined protocol must be followed in order to collect enough data in a timely manner to form a conclusion. Therefore, a heuristic has to be developed to guide the search for possible stable zero-bias peaks with a quantized height. In the experiments described in this manuscript, the procedure generally consists of the following steps. First, a fairly coarse search is conducted across gate voltages and magnetic field to locate possible zero-bias peaks. Once a candidate is found, its stability with respect to the control parameters, such as gate voltages and magnetic field, is checked. If a feature is deemed sufficiently stable, we determine whether the peak height approaches the quantized value for some parameter settings. If this is the case, further parameter sweeps are performed in an attempt to establish the extent of the parameter space over which the peak height remains close to $2e^2/h$. It should be noted that the same experimental tuning procedure could in principle be applied to search for plateaus at other conductance values, which would serve as an important cross-check of the assumption that a zero-bias peak with a constant height can only be achieved at $2e^2/h$. In practice, given time constraints, however, this cross-check was not performed systematically for conductance values other than $2e^2/h$.

Based on our experience with this experimental tuning heuristic, we propose a few conditions which, although perhaps not entirely sufficient, seem necessary before a feature can credibly be labeled as a plateau.

The first condition relates to the robustness of the zero-bias peak. The conductance spectroscopy signal must show a peak at zero bias for the entire range of the plateau-like region. This is in order to ensure that variations in the zero-bias conductance reflect a change in the peak height, rather than e.g. changes in the peak energy splitting while a parameter such as a gate voltage changes.

The second condition regards the stability of the peak height. The fluctuation of the peak height over the plateau-like region must not exceed 5% of the mean value. This 5% limit on the fluctuation allows to account for small variations due to experimental noise.

The third condition concerns the conductance behavior away from zero bias. The average derivative of the above-gap conductance with respect to the control parameter exceeds the average derivative of the zero-bias conductance by at least a factor 3 over the range of the plateau-like region. This to ensure that the peak height remains constant, while the control parameter ($V_{TG}$) is in fact changing the sample in a meaningful way (by changing the tunnel barrier height). In principle, this should apply simultaneously to all control parameters to some degree, although depending on the theoretical model, the expected stability with respect to e.g. the super gate voltage (i.e. chemical potential) can be much worse than the expected stability with respect to e.g. the tunnel gate voltage (i.e. coupling to the lead). Also, the precise quantitative relationships between the experimental tuning parameters (e.g. super gate voltage and tunnel gate voltage) and the theoretical parameters (e.g. chemical potential and tunneling amplitude) are at best known only qualitatively through numerical simulations since the precise electrostatic boundary conditions applying to the actual sample are never exactly known.

With these conditions in hand, we zoom in on the plateau-like region in Fig. 3, shown in Fig. E1. Based on the first condition, we can bound the plateau-like region on the left side at $V_{TG}$ = -7.93 V. On the right side, we bound the region by the significant charge jump at $V_{TG}$ = -7.6 V. Taking the average over this range for both the zero-bias conductance and $G_N$, we can compare the relative changes between the two, as well as the fluctuations. We see that while the zero-bias conductance generally stays within 5% of the average value, $G_N$ changes by almost 40% over the entire range. Additionally, while the zero-bias conductance fluctuates around the average with a small average derivative, $G_N$ follows an increasing trend. We could therefore state that our conditions have been met, and in this respect this feature can be considered as a plateau with reasonable stability.

In contrast, we take another look at the feature present in Fig. 5, with the plateau analysis shown in Fig. E2. In this case, we again use the first condition to bound the region on the left at $V_{TG}$ = -7.88 V, and a charge jump to bound the region on the right at $V_{TG}$ = -7.67 V. We see that in this case, the zero-bias conductance varies considerably more than 5% of the average value. Additionally, the average derivative of the zero-bias conductance actually exceeds the average derivative of $G_N$. Based on these considerations this feature does not qualify as a plateau according to our protocol.

The conditions we have set here are somewhat subjective. For example, if one widens or narrows the accepted range of the conductance fluctuations, varies the extent of the tuning parameter range, or demands a larger difference between the derivatives at zero bias and the above-gap conductance, one could reach different conclusions regarding specific data sets. As an example, using the plateau range defined by the vertical dotted green lines in Fig. 8c we would conclude that a plateau is present based on the conditions we have set. In Fig. 8d, the conductance varies too much to qualify the selected range as a plateau. However, one could choose to extend the selected range to the right until the second and third conditions are violated for Fig. 8c, or to crop the range slightly such that they are satisfied for Fig. 8d.

To give another example, if one cuts the considered region in Fig. E2 in half and only uses the data between $V_{TG}$ = -7.78 V and $V_{TG}$ = -7.67 V, we can find a "plateau" at 2.8 $e^2/h$ which satisfies our first and second conditions, but not the third, as the ratio of the derivatives is only 1.6. However, if the third condition is applied less strictly, and we only demand that the average derivative of the above-gap conductance exceeds that of the zero-bias conductance, this feature would pass the test.

It is not unlikely that given a sufficiently large data set, regions of parameter space can generically be found which satisfy the types of conditions we have set out. It is also apparent that the described tuning heuristic and our assignment protocol can influence whether or not a plateau can be found at a specific value. This analysis is therefore not intended to draw sharp distinctions, but rather to illustrate the uncertainties when it comes to a definition of a plateau in an experimental context. Such ambiguities exist in defining stability of conductance quantization in all topological systems, particularly in the early days of the experimental developments. Examples include the quantum spin Hall effect [14], the quantum anomalous Hall effect [15] and the 5/2 fractional quantum Hall plateau [16, 17]. A continued effort is needed to understand and improve the microscopic conditions for a quantized plateau with an acceptable experimental accuracy as well as the ramifications for the theoretical interpretation.

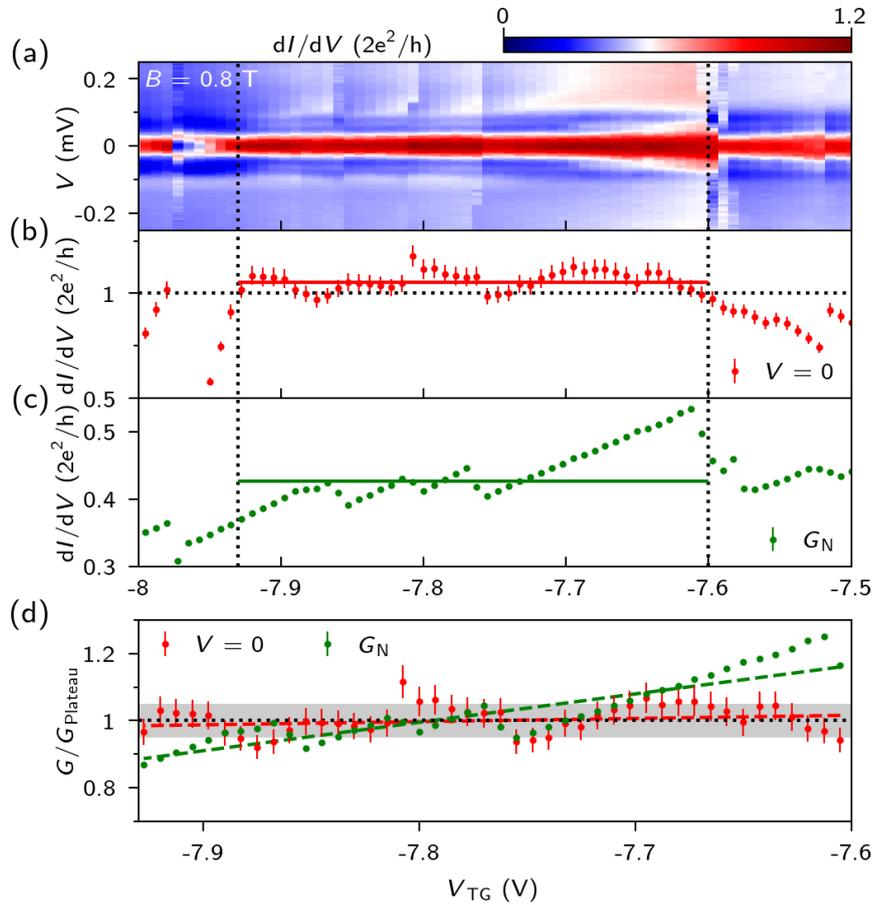

Figure E1. (a-c) Zoom in of Fig. 3a-c. The vertical black dotted lines indicate the candidate plateau-like region, i.e. the region under consideration for the test outlined in the text. The red and green horizontal lines in panels b and c indicate the average conductance in this region, which we call $G_{Plateau}$. (d) The relative conductance as a fraction of $G_{Plateau}$. The dashed lines are linear fits over the plateau region, indicating the average derivative for the zero-bias conductance (red) and $G_N$ (green). The gray shaded area shows a 5% window around the value of $G_{Plateau}$.

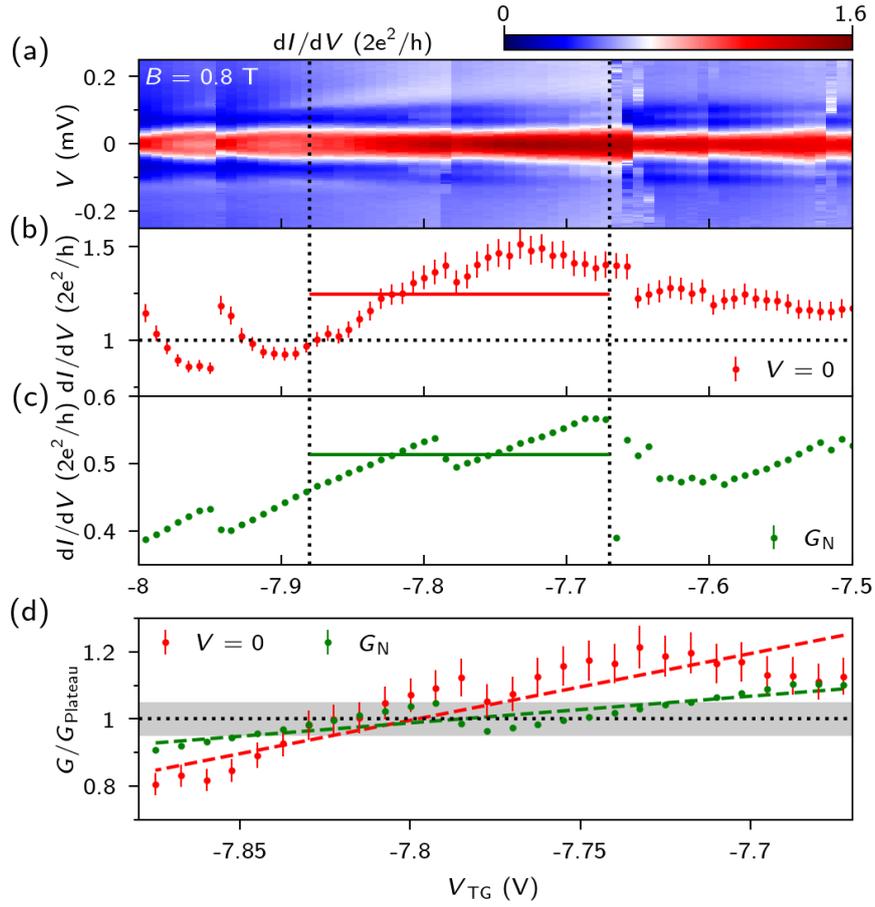

Figure E2. (a-c) Zoom in of Fig. 5a-c. The vertical black dotted lines indicate a candidate plateau-like region, which does not meet the stability condition. The red and green horizontal lines in panels b and c indicate the average conductance in this region, which we call $G_{Plateau}$. (d) The relative conductance as a fraction of $G_{Plateau}$. The dashed lines are linear fits over the plateau region, indicating the average derivative for the zero-bias conductance (red) and $G_N$ (green). The gray shaded area shows a 5% window around the value of $G_{Plateau}$.

# Appendix F. Measurement timeline

As mentioned, our devices tend to be relatively unstable, making it difficult to maintain consistent measurement conditions over time. Charging and discharging charge traps in the environment, either due to sudden jumps or from actively sweeping the gate voltage, combined with gate drifts on the scale of hours complicate the comparison of different data sets if they were measured sufficiently far apart. This distance can be measured either in time or in the amount of parameter space which was explored in between different measurements. In Table F1, we list the date and time each data set was taken, as well as the range over which various parameters had been explored since the previous entry in the table. When data sets are taken consecutively (labeled 'CS'), we expect the results to generally be consistent between them. However, if sufficiently time and/or parameter distance separates two data sets, we expect the settings cannot be compared directly.

| Data set | Device | Date | Time started | Time finished | Parameter range since previous entry | | | |
|---|---|---|---|---|---|---|---|---|
| | | | | | $V_{TG}$ (V) | $V_{SG}$ (V) | $V_{BG}$ (V) | $B$ (T) |
| Fig. 8b | B | 3-3 | 14:45 | 18:40 | - | - | - | - |
| Fig. 8c | B | 8-3 | 19:41 | 21:54 | -2.0 – -0.75 | -2.1 – 1.5 | -2.0 – -0.75 | 0 – 1.2 |
| Fig. 8d | B | 8-3 | 23:00 | 23:45 | CS | CS | CS | CS |
| Fig. H1 | B | 14-3 | 14:50 | 00:35 | -2.5 – -1.11 | -0.9 – 5.25 | -2.5 – -1.11 | 0 – 1.3 |
| Fig. 6c | A | 27-3 | 11:52 | 18:04 | - | - | - | - |
| Fig. 7a | A | 27-3 | 23:09 | 03:12 | CS | CS | Fixed | CS |
| Fig. 6a | A | 28-3 | 03:14 | 07:16 | CS | CS | Fixed | CS |
| Fig. 6b | A | 28-3 | 07:19 | 11:21 | CS | CS | Fixed | CS |
| Fig. 2b | A | 28-3 | 12:27 | 15:34 | -6.5 – -3.5 | -9.0 – -7.8 | Fixed | 0.6 – 0.9 |
| Fig. 7b | A | 28-3 | 15:35 | 18:29 | CS | CS | Fixed | CS |
| Fig. 2a | A | 29-3 | 01:03 | 02:45 | -9.0 – -6.0 | -6.5 – -5.5 | Fixed | 0 – 1.0 |
| Fig. 2c | A | 29-3 | 04:26 | 06:08 | -7.92 – -7.74 | Fixed | Fixed | 0 – 1.0 |
| Fig. 3 | A | 30-3 | 13:46 | 15:45 | -9.21 – -6.75 | -7.5 – -4.5 | Fixed | 0 – 1.0 |
| Fig. 5 | A | 31-3 | 20:07 | 22:36 | -8.25 – -7.5 | Fixed | Fixed | 0 – 1.0 |
| Fig. C1 | C | 7-4 | 14:48 | 23:20 | - | - | - | - |
| Fig. D1b | C | 7-4 | 23:21 | 08:38 | CS | CS | CS | CS |
| Fig. 9d | C | 18-4 | 15:23 | 18:52 | - | - | - | - |
| Fig. 9c | C | 18-4 | 19:13 | 20:43 | CS | CS | CS | CS |
| Fig. 9a | C | 18-4 | 23:47 | 02:11 | -0.1 – -0.03 | -1.5 – 1.5 | Fixed | Fixed |
| Fig. 9b | C | 23-4 | 19:23 | 20:18 | -0.1 – 0.1 | -1.5 – 1.5 | -22.5 – 0 | -0.1 – 1.1 |

Table F1. Measurement timeline of all the data sets included in this manuscript. For the parameter changes, we describe the range over which these parameters were since the previous entry. We indicate consecutive measurements by 'CS'. If the parameter remains unaltered until the measurement is initiated it is labeled 'Fixed'. All data sets were taken in 2017.

# Appendix G. Comparison of Lorentzian fits with FWHM analysis

In Fig. 4, we show the peak height and width as extracted from the data in Fig. 3 via a fit to a Lorentzian line shape thermally broadened by 20 mK:

$$G(V, T = 20 \text{ mK}) = \int d\epsilon \, G(\epsilon, 0) \frac{d}{d\epsilon} f(eV - \epsilon, T = 20 \text{ mK}), \qquad G(E, 0) = G_0 \frac{\Gamma^2}{\Gamma^2 + E^2}$$

Here, the peak height is given by $G_0$, and the peak width is $2\Gamma$. An alternative way to determine these parameters is to take the conductance value at $V = 0$ as the height and take the full width at half maximum (FWHM) for the width. This method is less dependent on the exact peak shape, which can be advantageous when analyzing peaks which are broadened beyond the Lorentzian regime [13].

Fig. G1a shows an example of this procedure, with the results for the full data set used in Fig. 4 shown in Fig. G1b. Comparing the results for the height (Fig. G1c) and width (Fig. G1d), we see that although the methods show some quantitative differences, especially for the broader peaks, they produce the same qualitative trends. The peak width has a positive correlation with $G_N$, while the peak height shows no clear correlation with either the peak width or $G_N$ (as shown in more detail Fig. G3a).

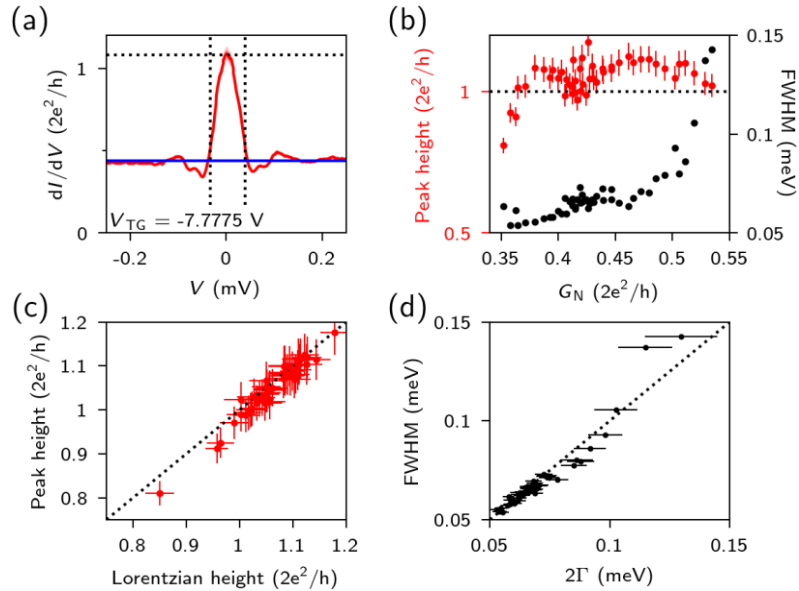

Figure G1. (a) Example of the peak height and width analysis used in panel (b). The height is determined by $dI/dV$ at $V = 0$ (dotted horizontal line). The FWHM is then determined by finding the values of $V$ where the conductance equals half the peak conductance (vertical dotted lines). The extracted value of $G_N$ is shown by the blue line. (b) Peak height (red circles) and FWHM (black circles) as a function of $G_N$. (c) Comparison of the peak height found by taking the conductance value at $V = 0$ with the height extracted from the Lorentzian fit. The dotted line indicates the points where the two values are equal. (d) Comparison of the FWHM with the peak width extracted from the Lorentzian fit ($2\Gamma$). The dotted line indicates the points where the two values are equal.

In Fig. G2 we show the same analysis performed on the data set reported in Fig. 9b. Again, we see that the Lorentzian and FWHM analysis give qualitatively similar results. However, in this case the peak height has a positive correlation with $G_N$, and a negative correlation with the peak width, indicating that this peak is most likely due to a level crossing (see Fig. G3b).

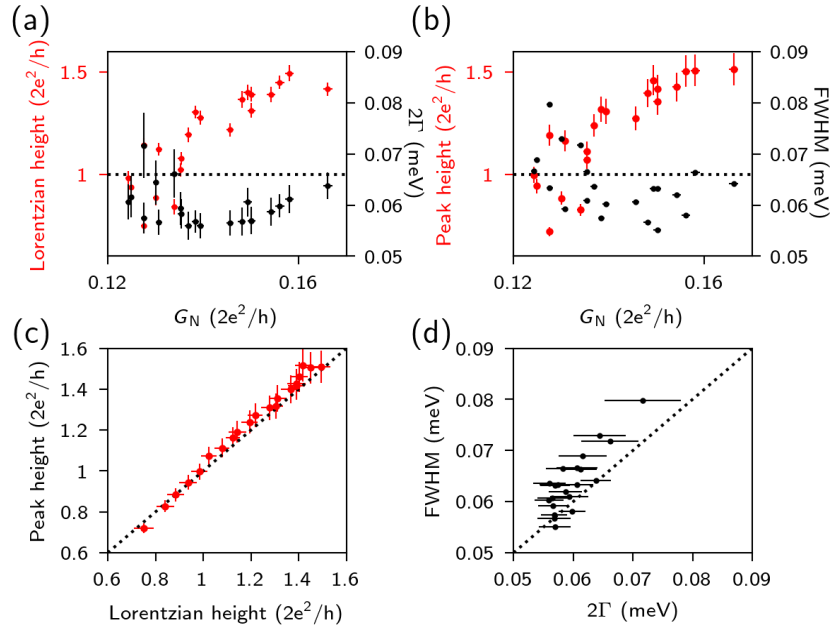

Figure G2. (a) Results of the Lorentzian fit of the peaks in Fig. 9b in the range $V_{BG}$ = -11.9 V to $V_{BG}$ = -10.9 V. (b) Results of the FWHM analysis on the same data set (c) Comparison of the peak height found by taking the conductance value at $V$ = 0 with the height extracted from the Lorentzian fit. The dotted line indicates the points where the two values are equal. (d) Comparison of the FWHM with the peak width extracted from the Lorentzian fit ($2\Gamma$). The dotted line indicates the points where the two values are equal.

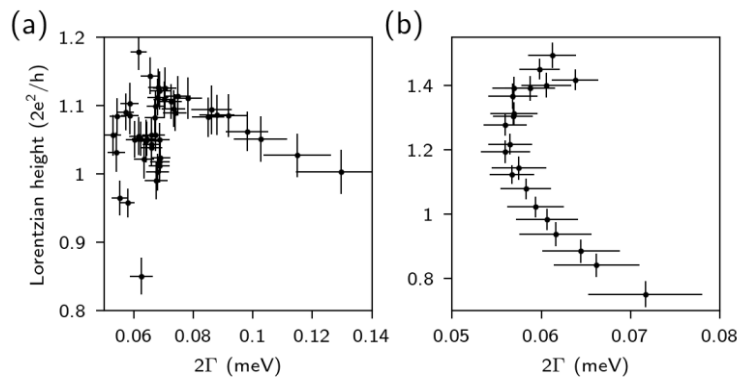

Figure G3. (a) Peak height as obtained by Lorentzian fit as a function of the peak width $2\Gamma$ for the data presented in Fig. 3. The Pearson's correlation coefficient is 0.009, indicating there is little to no linear correlation between the two variables. The decreasing trend in the peak height for the widest peaks could be indicative of the onset of peak splitting. (b) Peak height as a function of peak width for the data presented in Fig. 9b. The correlation coefficient is -0.562, indicating a negative correlation between the peak height and the peak width.

## Appendix H. Temperature dependence of the ZBP height

At $T = 0$, the conductance through a Majorana zero mode should be quantized at $2e^2/h$ independent of the tunnel coupling to the lead. While the peak height is constant, the peak width is determined by this tunnel coupling. At finite temperature, thermal broadening of the Fermi edge in the metallic lead gives rise to an increase in the peak width as well as a decrease in the peak height, with the effect size depending on the ratio of the tunnel and thermal broadening [13]. In Fig. H1a, we show the temperature dependence of the ZBP reported in Fig. 8 of the main text. From the line traces in Fig. H1b, we see that as the temperature is increased, the peak height decreases while the peak width increases.

For any discrete resonance coupled to a reservoir, the conductance at temperature $T$ can be calculated from the conductance at $T = 0$ via a convolution with the derivative of the Fermi-Dirac distribution $f$:

$$G(V,T) = \int d\epsilon\, G(\epsilon, 0) \frac{d}{d\epsilon} f(eV - \epsilon, T)$$

For a given peak shape, this can be converted to a universal scaling function which only depends on the ratio of the tunnel coupling and the temperature [18].

To simulate the effect of the increasing temperature, we use the data at base temperature ($T = 20$ mK) to approximate $G(V,0)$, and calculate the expected conductance at higher temperatures. Fig. H1c shows the measured conductance at a few selected temperatures in black, with the simulated trace shown in red. We see that there is a good agreement between the two, indicating that the decrease in peak height can be attributed to the increase in temperature through the increase of the thermal broadening. This is further demonstrated in Fig. H2. For low temperatures, the peak width is mostly determined by the tunnel broadening, and the peak height approaches a saturation of 2.3 $e^2/h$ as the temperature is decreased towards 0. When the temperature is increased, the relative importance of the thermal broadening mechanism increases, leading to a decrease in the peak height.

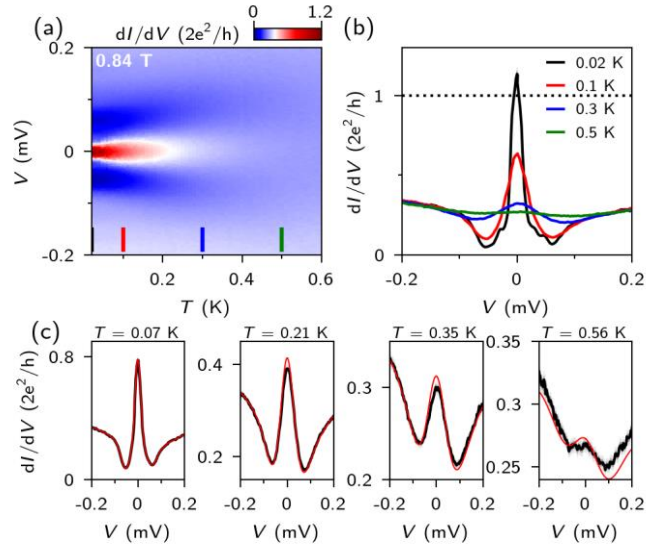

Figure H1. Temperature dependence of the ZBP height. (a) Differential conductance d$I$/d$V$ as a function of voltage $V$ and temperature $T$ at $B$ = 0.84 T. $V_{SG}$ = -0.39 V, $V_{TG}$ = $V_{BG}$ = -1.255 V. (b) Line traces from panel (a), illustrating the decrease in the peak height as the temperature is increased. (c) Examples of the zero-bias peak height for different temperatures. Experimental data is shown as the black curves, while the red curves are obtained by taking the conductance at base temperature (20 mK) and performing a convolution with the derivative of the Fermi distribution at the given temperature. Data was obtained from device B.

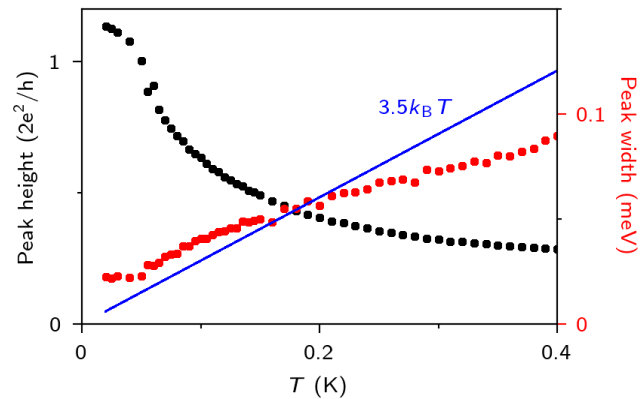

Figure H2. Extracted ZBP height (black) and width (red) as a function of temperature. As the temperature increases, thermal broadening starts to dominate over tunnel broadening in determining the peak width, causing the ZBP height to decrease. The blue line indicates the energy scale associated with the thermal broadening, $3.5k_BT$.